\documentclass[11pt, a4paper]{article}
\pdfoutput=1

\usepackage{jheppub}
\usepackage[utf8]{inputenc}
\usepackage{mathtools}
\usepackage{latexsym}
\usepackage{appendix}
\usepackage{mathrsfs}
\usepackage{graphicx}
\usepackage{microtype}
\usepackage{feynmp-auto}
\usepackage{float} 
\usepackage[caption = false]{subfig} 

\usepackage{shuffle}

\def\ab{\left[ \begin{smallmatrix} {\beta}\\
      {\alpha} \end{smallmatrix} \right]}

\def\be{\begin{equation}}
\def\ee{\end{equation}}
\def\nn{\nonumber}
\def\A{\mathcal{A}}
\def\C{\mathbb{C}}
\def\CP{\mathbb{CP}}
\def\Q{\mathbb{Q}}
\def\R{\mathbb{R}}
\def\Z{\mathbb{Z}}
\def\I{\mathcal{I}}
\def\J{\mathcal{J}}
\def\K{\mathcal{K}}
\def\M{\mathcal{M}}
\def\bdry{\partial}

\renewcommand{\Im}{\operatorname{Im}}

\DeclareMathOperator{\SL}{SL}

\title{Monodromy relations from twisted homology}

\author[a]{Eduardo Casali,}\emailAdd{ecasali@ucdavis.edu}
\affiliation[{a}]{Center for Quantum Mathematics and Physics (QMAP) and\\
Department of Physics, University of California, Davis, CA 95616 USA}

\author[{b,c,d}]{Sebastian Mizera,}\emailAdd{smizera@ias.edu}
\affiliation[{b}]{Perimeter Institute for Theoretical Physics, Waterloo, ON N2L 2Y5, Canada}
\affiliation[{c}]{Department of Physics \& Astronomy, University of Waterloo, Waterloo, ON N2L 3G1, Canada}
\affiliation[{d}]{Institute for Advanced Study, Einstein Drive, Princeton, NJ 08540, USA}

\author[{e}]{Piotr Tourkine}\emailAdd{piotr.tourkine@lpthe.jussieu.fr}
\affiliation[{e}]{LPTHE, CNRS \& Sorbonne Universit\'e, 4 Place Jussieu, 75005 Paris, France}

\abstract{We reformulate the monodromy relations of open-string scattering amplitudes as boundary terms of twisted homologies on the configuration spaces of Riemann surfaces of arbitrary genus.
  This allows us to write explicit linear relations involving loop integrands of open-string theories for any number of external particles and, for the first time, to arbitrary genus.
  In the non-planar sector, these relations contain seemingly unphysical contributions, which we argue clarify mismatches in previous literature.
  The text is mostly self-contained and presents a concise introduction to twisted homologies.
  As a result of this powerful formulation, we can propose estimates on the number of independent loop integrands based on Euler characteristics of the relevant configuration spaces, leading to a higher-genus generalization of the famous $(n-3)!$ result at genus zero.
}

\begin{document}
\maketitle

\section{Introduction}
\label{sec:introduction}

Gauge theory scattering amplitudes in open string theory are computed as integrals over middle-dimensional cycles on the moduli space of Riemann surfaces with punctures inserted on its boundaries. The color structure of a given amplitude is encoded in the choice of the integration contour, corresponding to the order in which vertex operators are inserted, while the integrand captures the information about the matter content. It has long been appreciated that amplitudes with different color and kinematics structures are not mutually independent, see, e.g., \cite{Plahte:1970wy} and more recently~\cite{BjerrumBohr:2009rd,BjerrumBohr:2010zs,Stieberger:2009hq}. In order to minimize the amount of computations it is thus beneficial to find bases of string integrals as well as ways of relating arbitrary amplitudes to such bases.

Another motivation for constructing these bases comes from the celebrated Kawai--Lewellen--Tye (KLT) \cite{Kawai:1985xq} relations, which express tree-level closed string amplitudes as a quadratic sum over bases of tree-level open string amplitudes ``glued'' together by a matrix of external kinematics invariants. Their incarnation in field theory, known as the color-kinematics duality~\cite{Bern:2008qj,Bern:2010ue}, has had tremendous impact in the development of modern methods to compute scattering amplitudes: we refer the interested reader to the recent review \cite{Bern:2019prr} and references therein. A generalization to loop-level in string theory is still missing and would be highly desirable.
One of the reasons why a generalization of these relations to higher genus has remained elusive might be that the simple contour-deformation arguments originally used to prove the tree-level result \cite{Kawai:1985xq} do not seem to generalize well on higher-genus surfaces.
Fortuitously, it has been recently shown~\cite{Mizera:2017cqs} that KLT relations have a geometric interpretation in terms of intersection numbers of bases of integration cycles on the moduli space of the $n$-punctured Riemann spheres, ${\cal M}_{0,n}$.
In this language KLT relations are understood simply as an insertion of a resolution of identity in the space of integration cycles.

Thus studying relations between amplitudes and constructing their bases amounts to characterizing these spaces. These spaces of integration cycles can be identified as homology groups with coefficients in a \emph{local system}, often called \emph{twisted homology groups}. A local system is a representation of the fundamental group that describes monodromy properties of moduli spaces when punctures encircle one another or travel around handles of a Riemann surface. Twisted homologies should be thought of as being ``non-perturbative'' in $\alpha'$, in the sense that they allow for computations exact in the inverse string tension $\alpha'$, in contrast with homology groups with constant coefficients (such as $\Q$), which are typically employed in ``perturbative'' calculations around the $\alpha' \to 0$ limit in terms of iterated integrals, see, e.g., \cite{Brown:2009qja,brown2011multiple,Brown:2018omk}.

At genus zero twisted geometries have been used to identify the KLT matrix in terms of bilinear pairings leading to new ways of calculating it, either through intersection numbers on the twisted homology~\cite{MANA:MANA19941660122} or algebraically in the twisted \emph{cohomology}~\cite{zbMATH03996010,cho1995}. Similar techniques have also been used to prove the existence of bases of integration cycles and integrands together with ways of projecting arbitrary amplitudes onto them, as well as provide means for regularization and analytic continuation of string integrals \cite{Mizera:2017cqs,Mizera:2019gea}, see also \cite{Hanson:2006zc,Gaiotto:2013rk,Witten:2013pra,Brown:2019wna} for earlier work. The success of this reinterpretation of tree-level string amplitudes begs the question of what are their higher-genus analogues, and if they are similarly powerful. In particular, they might give a natural way of generalizing KLT relations to higher-genus moduli spaces.

\begin{figure}[!h]\label{g_cut_intro}
  \centering
    \subfloat[Surface before cutting]{\includegraphics[scale=1]{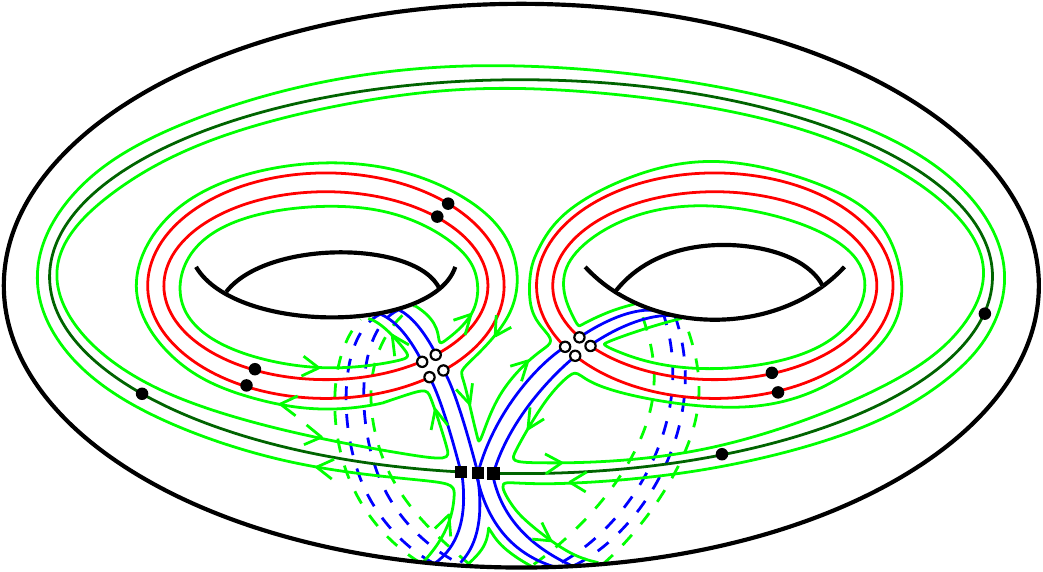}\label{fig:g_cut_intro_1}}\\
  \subfloat[After cutting along the A cycles. In figure~\ref{fig:gcut_2} we further cut open along the B cycles.]{\includegraphics[scale=1]{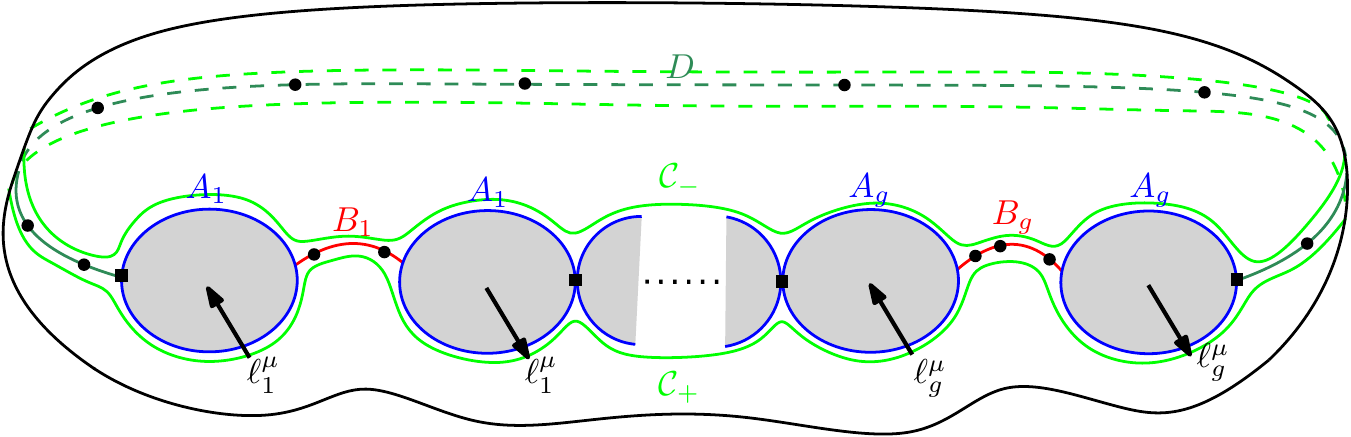}\label{fig:g_cut_intro_2}}
 	\caption{Genus-$g$ Riemann surface sliced along the $A_I$-cycles (blue). Loop momenta $\ell_I^\mu$ flow through the two pairs of resulting disks for each $I=1,2,\ldots,g$ (the point denoted with a black box is identified). The resulting Riemann surface can be understood as two copies of open-string worldsheets glued together along $B_I$-cycles (red) and $D$ (dark green), on which punctures are placed. Monodromy relations are a consequence of the fact that the contours ${\cal C}_{\pm}$ (light green) can each be contracted to a point.}
      \end{figure}

The aim of the present paper is to pave a way towards this higher-genus understanding. The first question one should ask is, keeping all but one particle fixed, what are the bases of integration cycles in which higher-genus amplitudes are computed? This question can be addressed using twisted homology on a \emph{single} Riemann surface with punctures, as opposed to the much richer twisted homology on the moduli space ${\cal M}_{g,n}$, which is currently not understood. As a matter of fact, there are various ways for setting up the problem even for a single surface. In our setup we use chiral splitting \cite{Verlinde:1986kw,DHoker:1988pdl,DHoker:1989cxq} for open-string amplitudes, which allows for writing the amplitude as a holomorphic integral. At the same time it introduces loop momenta $\ell_I^\mu$ assigned to each $A_I$-cycle of the compact genus-$g$ Riemann surface with $I=1,2,\ldots,g$. In this way we can work on the surface sliced along those $A_I$-cycles which now has a topology of a disk. Integration cycles relevant to open-string amplitudes are those connecting two punctures, as in Figure~\ref{g_cut_intro}.

We then formulate twisted homology on such a cut Riemann surface keeping loop momenta $\ell_I^\mu$ as well as surface moduli $\Omega_{IJ}$ fixed. Therefore all statements we make are about loop \emph{integrands} for open strings. The introduction of loop momenta also helps in the interpretation of our results in the field-theory limit.

Key identities used to deduce relations between different color-ordered amplitudes at genus zero are the \emph{monodromy relations} introduced by Plahte \cite{Plahte:1970wy} and later studied in \cite{BjerrumBohr:2009rd,Stieberger:2009hq}.
From the perspective of twisted homology, they become simply a statement of the vanishing of terms which are a total boundary under the twisted boundary operator \cite{10.1093/qmath/38.4.385,Mizera:2019gea}, that is, contours that can be deformed to a point. Using this framework one can prove that the dimension of twisted homology of ${\cal M}_{0,n}$ is $(n{-}3)!$  \cite{aomoto1987gauss} (in the physics literature this result first appeared in \cite{BjerrumBohr:2009rd,Stieberger:2009hq}), which has a geometric interpretation in terms of the Euler characteristic of ${\cal M}_{0,n}$.

Our setup essentially trivializes finding monodromy relations for higher-genus surfaces. This allows us to write down explicit identities for all planar open-string integrands with particle inserted on any boundary to arbitrary genus $g$,\footnote{In order to make statements about all-genus results, throughout this work we assume that string-theory amplitudes can be consistently formulated on bosonic moduli spaces ${\cal M}_{g,n}$, as opposed to supermoduli spaces, at high-enough genus. See \cite{Donagi:2013dua,Sen:2015hia} for discussions of this problem. Alternatively, the pure spinor superstring proposes such a formulation \cite{Berkovits:2000fe,Berkovits:2006vi}.}
arbitrary number of punctures $n$, all orders in $\alpha'$, and any external string state. 
For the correct counting of the number of independent cycles it is important to realize that there are always \emph{two} monodromy relations for each surface, obtained from contracting the contours ${\cal C}_{+}$ and ${\cal C}_{-}$ in Figure~\ref{g_cut_intro}. These two relations explain heuristically the reduction from the naive $(n-1)!$ to $(n-3)!$ distinct tree-level amplitudes. If one builds up recursively the amplitude by adding punctures, at each step there are two fewer independent amplitudes resulting in the previous reduction. 
The story is sometimes presented differently in the literature, where complex conjugation, or real and imaginary parts, are invoked. While this is effectively identical to using these two relations we believe that the problem is better formulated in those terms.

At low genus ($g=1,2,3$), similar monodromy relations were recently written down in \cite{Tourkine:2016bak,Hohenegger:2017kqy,Ochirov:2017jby,Tourkine:2019ukp} and checked to the first couple of orders in $\alpha'$. These results were a source of a few discrepancies, which are resolved here. We extend to all genus the observation of \cite{Hohenegger:2017kqy} that certain ``unphysical'' cycles (those that do not appear naturally in string perturbation theory) contribute to monodromy relations. A consequence of this fact is that a basis of twisted homology \emph{cannot} be constructed purely out of ``physical'' cycles contributing to the open-string amplitudes.

We claim the new monodromy relations are all the \textit{generic} relations of open-string integrands obtained by moving one particle at a time, that is, they are valid for a generic open-string loop integrands. This leads us to a conjecture on the size of the basis to all loops, based on Euler characteristic computations, giving an upper bound on the number of independent string loop integrands.

Lastly, we hope that our construction of twisted homology for higher-genus surfaces and their moduli spaces will be relevant to the study of Feynman multi-loop integrals (where the role of $\alpha'$ is taken by the dimensional-regularization parameter $\varepsilon$), given their recent appearance in this context, see, e.g., \cite{Broedel:2018qkq,Vanhove:2018mto,Bogner:2019lfa} and references therein, as well as \cite{Mastrolia:2018uzb} for a related setup.

This paper is organized as follows. Section~\ref{sec:prelim} presents a didactic introduction to local systems, twisted homology, and their relation to tree-level string amplitudes. It also contains a derivation of the tree-level monodromy relations in the context of twisted homology. This section ends with a generalization to higher-genus surfaces.
Section~\ref{sec:genus_one} gives a detailed application of the tools introduced in the previous section to genus-one string amplitudes; here we re-derive the genus-one monodromy relations and compare it to other relations in the literature. In Section~\ref{sec:higher_g} we generalize the local system to higher-genus and use it to explicitly derive monodromy relations to all genus. Lastly, in Section~\ref{sec:discussion} we discuss the counting of independent integrands, comment on the field-theory limit of our monodromy relations and their connection to Bern--Carrasco--Johansson (BCJ) \cite{Bern:2008qj} representation of field theory loop integrands, and discuss further steps towards loop-level KLT.

\section{Preliminaries}
\label{sec:prelim}

This section serves as an introduction to the framework used in the paper. It starts with a pedagogical example which illustrates how the language of local systems simplifies computations involving multi-valued functions.
We then review genus-zero string theory amplitudes in the context of twisted homologies and the interpretation of their monodromy relations in this language. After that, we introduce the set-up which allow us to generalize the genus-zero computation to higher genera.

\subsection{From multi-sheeted surfaces to local systems}

\begin{figure}
	\centering
	\subfloat[Path $\gamma$ crossing a branch cut at points $a$ and $b$.]{\includegraphics{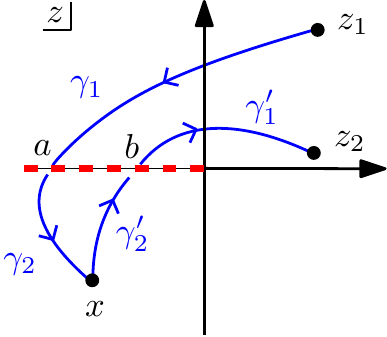}\label{fig:equiv1}}\quad
	\subfloat[Path $\gamma$ on the multi-sheeted surface $\widehat{X}$.]{\includegraphics[scale=0.8]{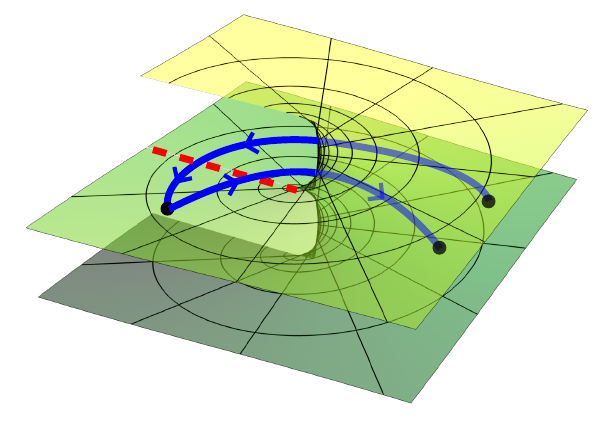}\label{fig:equiv2}}\quad
	\subfloat[Equivalence to picking up a monodromy factor.]{\includegraphics{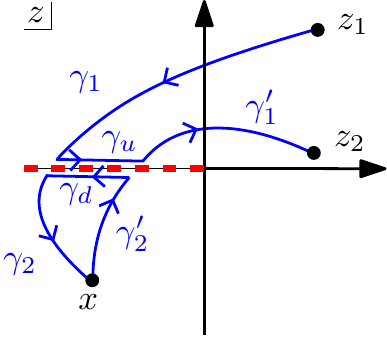}\label{fig:equiv3}}
	\caption{Translation between multi-valued integrands on the covering space $\widehat{X}$ and paths with local coefficients on the base space $X$, which gives an illustration of what is meant by ``picking up a phase when crossing a branch cut''.}
	\label{fig:equiv}
\end{figure}

Let us consider a function $f(z)$ with a branch cut taken along the negative real axis of a complex plane. For instance,
\begin{equation}
f(z)=z^s,\label{eq:f-def}
\end{equation}
where $s$ is a non-integer positive real number.
This function defines a multi-sheeted Riemann surface branched at the origin and infinity of the type illustrated in Figure~\ref{fig:equiv2}.
Let us consider an integral of $f(z)$ along a path $\gamma$ crossing its branch cut (which we take to lie on the negative real axis), say at two points, as shown in Figure~\ref{fig:equiv1} and \ref{fig:equiv2}. Since the integrand is multi-valued, rather than the original punctured complex plane $X=\C-\{0\}$, the integration is defined on its universal cover $\widehat{X}$, which is the surface shown in Figure~\ref{fig:equiv2}. In order to simplify this problem we would like to learn how to give a meaning to
\begin{equation}
  \label{eq:path-cross}
  I := \int_\gamma f(z)
\end{equation}
as an integral directly on the original space $X$. In the example at hand this might not seem like a significant simplification, however it will give us the understanding of how to deal with more general cases where the covering space might be much more complicated.\footnote{Strictly speaking, instead of the universal cover $\widehat{X}$ for our purposes it is sufficient to consider the maximal Abelian cover of $X$ (its covering group is the first homology group $H_1(X,\Z)$, which is the Abelianization of the fundamental group $\pi_1(X)$). This will be reflected in the definition of the local system reviewed in Section~\ref{sec:review}.}

As the first step, we split $\gamma$ into a collection of paths $\gamma_1, \gamma_2, \gamma_2', \gamma_1'$ which do not cross the branch cut, see Figure~\ref{fig:equiv1} (the points $a$ and $b$ can be included in either of these paths, this choice does not affect the argument).
In fact, we might as well complete this set by including $\gamma_u$ and $\gamma_d$, which lie on the two sides of the branch cut, see Figure~\ref{fig:equiv3}. If we were on $\widehat{X}$, integrations over the two of them ought to cancel out.

In order to define integrals over each path uniquely on $X$, we can select a ``principal branch'' of $f(z)$ and agree to take all integrals to start at this specific branch. In these conventions we have:
\begin{equation}\label{eq:gamma-u-d}
\int_{\gamma_u} f(z) = -e^{2\pi i s}\!\! \int_{\gamma_d}f(z).
\end{equation}
Here the minus sign accounts for the opposite orientations, while the factor of $e^{2\pi i s}$ on the right-hand side compensates for the fact that the integrand of $\int_{\gamma_d}$ starts on the principal branch instead of being analytically-continued from that of $\int_{\gamma_u}$ by crossing a branch cut.

This exercise suggests a general rule, in which the original integration along $\gamma$ on the covering space $\widehat{X}$ translates to $X$ as ``picking up'' a monodromy factor of $e^{2 \pi i s}$ each time the cut is crossed downwards, and likewise a factor of $e^{-2 \pi i s}$ if it is crossed in the upwards direction. In this way, the integral \eqref{eq:path-cross} can be written on $X$ as a sum
\begin{equation}
  \label{eq:twisted-cycle}
  I = \left(\int_{\gamma_1}
  +\; e^{2\pi i s}\!\!\int_{\gamma_2}
  +\; e^{2\pi i s}\!\!\int_{\gamma_2'}
  +\; \int_{\gamma_1'} \right) f(z),
\end{equation}
according to the above rules.

As a sanity check let us confirm that this definition is homotopy-independent. In particular, it should not depend on the position of the points $a,b,x$, so the integral must be equal to one along a straight line connecting $z_1$ and $z_2$. Indeed, combining \eqref{eq:gamma-u-d} and \eqref{eq:twisted-cycle} we find
\begin{equation}
I = \left(
\int_{\gamma_1}
+\; \int_{\gamma_u}
+\; \int_{\gamma_1'}
\right) f(z)
+
e^{2\pi i s} \left(
\int_{\gamma_2'}
+\; \int_{\gamma_d}
+\; \int_{\gamma_2}
\right) f(z).
\end{equation}
Here the three integrals in the first bracket become $\int_{z_1}^{z_2}$ by contour deformation, while those in the second bracket sum to zero by Cauchy's theorem since the contour $\gamma_2' + \gamma_d + \gamma_2$ is contractible to a point and $f(z)$ does not have any poles on $X$.

This simple example illustrates the setup that we will be working with; the arbitrary choice of branch cuts and their corresponding monodromy factors describes a so-called \emph{local system} on $X$. 
As in \eqref{eq:twisted-cycle}, each path is allowed to have a \emph{local coefficient}, such as $1$ or $e^{2\pi i s}$, which provides an additional piece of information describing this path.
One should think of these coefficients as an effect of ``collapsing'' all branches of $\widehat{X}$ onto the original space $X$, which is also a book-keeping device that remembers which cuts have been crossed and how. For the reader familiar with defining integrals along cuts with contour deformation, or even more sophisticated concepts such as Pochhammer contours, we can only emphasize that this way of looking at the problem saves a huge amount of combinatorial complexity in analyzing integrals of branched functions. For a longer pedagogical review see Appendix~A of \cite{Mizera:2019gea}.

The appropriate mathematical language for describing local systems and the associated integration theory is that of twisted homologies and cohomologies, (we sometime write below ``(co)homologies'') introduced below in the context of tree-level string amplitudes.

\subsection{Review: twisted homology at genus zero}
\label{sec:review}

The main idea behind using twisted (co)homology to describe open-string amplitudes relies on the universality of the so-called Koba--Nielsen factor, introduced in eq.~\eqref{tree_open} below. As the first step we analytically continue the integrand from open-string to closed-string moduli space, in which positions of punctures (insertions of vertex operators) are promoted from real to complex coordinates. This introduces multi-valuedness in the Koba--Nielsen factor as a function of puncture coordinates.
The twisted (co)homology then can be thought as a (co)homology theory which keeps track of the multi-valuedness of this function. Heuristically, and element of a twisted (co)homology group is given by a pair: and element of the usual (co)homology and a number which is a monodromy of the Koba--Nielsen factor.

Here we will only work explicitly with the twisted \emph{homology} theory,\footnote{One could also consider similar computations in twisted \textit{cohomology} as an extension of the genus-zero case studied in \cite{Mizera:2017cqs,Mizera:2017rqa,Mizera:2019gea}.} interchangeably called homology with coefficients in a \emph{local system} or with \emph{local coefficients}~\cite{10.2307/1969099,aomoto1977structure,aomoto2011theory}.
For an introduction geared towards physical applications  see~\cite{Mizera:2017cqs,Mizera:2019gea}. Below we give a working definition of local systems and twisted homology that is needed to study the monodromy relations.

Let us review the simplest case relevant for string-theory amplitudes, a tree-level open-string amplitude, and introduce the ideas of twisted homology using this example. Consider the tree-level open-string amplitude $\A(\alpha)$ with color ordering $\alpha$, we fix the $\SL(2,\mathbb{C})$ redundancy by setting $\{z_{n-2},z_{n-1},z_{n}\}$ to some particular values. This amplitude can be represented by
\begin{equation}\label{tree_open}
\A(\alpha) := \int_{\Delta(\alpha)} \prod_{j=1}^{n-3}d z_j\!\! \prod_{1 \leq j<l\leq n} \!\!\! |z_l{-}z_j|^{\alpha' k_j\cdot k_l}\, \varphi(z_j).
\end{equation}
In this expression, $\alpha'$ is the inverse string tension, $\varphi(z_j)$ is a theory-dependent, rational function of the location of the punctures $z_j$, which can also depend on external variables such as coupling constants, momenta $k_j^\mu$, or polarization vectors $\varepsilon_j^\mu$ of the $j$-th particle. Here $\Delta(\alpha)$ is a top-dimensional cycle on the moduli space of disks with $n$ punctures on the boundary $\M_{0,n}(\R)$, given by
\begin{equation}\label{Delta-alpha}
\Delta(\alpha) := \{ z_{\alpha(n)} < z_{\alpha(1)} < z_{\alpha(2)} < \dots < z_{\alpha(n)}\},
\end{equation}
where the first inequality is understood in cyclic sense since the boundary of a disk is $\mathbb{RP}^1$. The color ordering $\alpha$ enters only through the choice of integration cycle, as well as implicitly in the Koba--Nielsen factor $\prod_{j<l} |z_l {-} z_j|^{\alpha' k_j \cdot k_l}$, which involves absolute values.

\begin{figure}[t]
	\centering
	\includegraphics[scale=1]{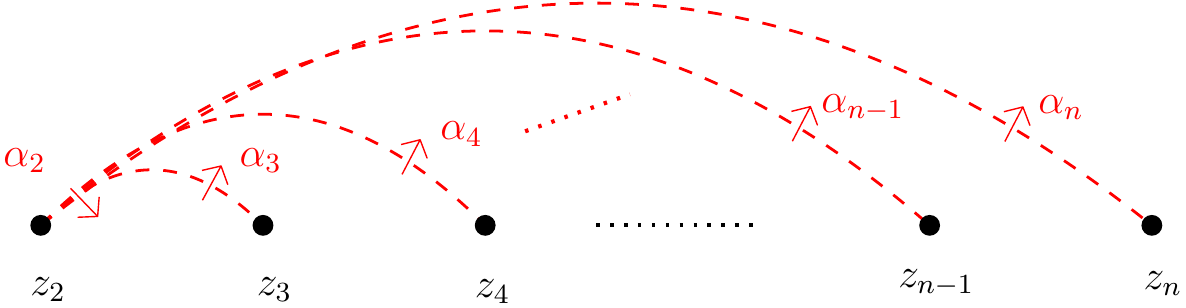}
	\caption{Choice of branch cuts for the loading $T_0(z_1)$ with monodromies $\alpha_j=e^{-2\pi i \alpha'k_1\cdot k_j}$. Note that $\alpha_2 = \prod_{j=3}^{n} \alpha_j^{-1} = e^{-2\pi i \alpha' k_1 {\cdot} k_2}$ is not independent of the other ones.}
	\label{branches_g0}
\end{figure}

Here we make use of the canonical embedding of the open-string moduli space into the closed-string moduli space, $\M_{0,n}(\R) {\subset} \M_{0,n}$, which gives a way of analytically continuing the integrand of \eqref{tree_open} to the complex space $\M_{0,n}$. Different color orderings on the open-string amplitudes are now understood as distinct choices of middle-dimensional contours in the closed-string moduli space.

We will start by considering monodromies of the integrand of \eqref{tree_open} as a function of $z_1$, keeping all other punctures $z_{j\neq 1}$ fixed. In more precise mathematical terms, this corresponds to working on a fibre $\Sigma_{0,n-1}$ of the forgetful map $\mathcal{M}_{0,n} \to \M_{0,n-1}$, which is equal to the Riemann sphere
with points $n{-}1$ removed, $\Sigma_{0,n-1} := \Sigma_0 {-} \{z_2, \dots z_n\}$.
In this case, the relevant analytic-continued part of the integrand is 
\begin{equation}\label{loading_ex}
 T_{0}(z_1) := \prod_{j=2}^n (z_j-z_1)^{k_1\cdot k_j}
\end{equation}
where in the above (and in the rest of this paper) we omit $\alpha'$ to make formulas more transparent. Here $T_0(z_1)$ is a multi-valued function defined on a cover of $\Sigma_{0,n-1}$, so we make an assignment of branch cuts as shown in Figure~\ref{branches_g0}, and corresponding monodromy factors
\begin{equation}
  \label{eq:alpha-i-def}
  \alpha_j := \exp(-2i\pi k_1{\cdot} k_j) \;\in\; \mathbb{C}^*\,.
\end{equation}
Note that there are $n{-}2$ independent branch cuts, which is the same as the number of independent (singular) one-cycles on an sphere with $n{-}1$ punctures.

We use the same Figure~\ref{branches_g0} to define a \textit{local system} $\mathcal{L}_0$ on $\Sigma_{0,n-1}$. It consists of an Abelian representation of the fundamental group of $\Sigma_{0,n-1}$,
\be
\pi_1(\Sigma_{0,n-1})\rightarrow \mathbb{C}^\ast,
\ee
which assigns a non-zero complex number $\alpha_j \in \C^\ast := \C - \{0\}$, called a \emph{monodromy}, to each loop on the surface (contractible loops have a unit coefficient). This defines a flat line bundle $\mathcal{L}_0$ with $\mathbb{C}^\ast$ fibres.
Because it is flat, its parallel transport around a contractible loop returns a section to itself, but when transported around a non-contractible loop it can get multiplied by a non-zero number (an element of the linear endomorphism $\mathrm{End}(\mathcal{L}_0)$).

This behavior captures monodromies of the local system, which are the non-zero complex numbers that multiply a section transported along a non-trivial cycle. In this context we interpret Figure~\ref{branches_g0} as providing the data for a local system, where the monodromies $\alpha_j$ are considered as elements of $\text{End}(\mathcal{L}_0)$ and combinations of them give the required representation of $\pi_1(\Sigma_{0,n-1})$.
In short: whenever a section of $\mathcal{L}_0$ is transported by a path that crosses a dashed red line, it gets multiplied by the respective $\alpha_j$. This definition of the local system depends on a choice of basis of the first singular homology group of the surface, $H_1(\Sigma_{0,n-1},\Z)$. At genus zero we can take this basis to coincide with the branch cuts of the function~\eqref{loading_ex} given in Figure~\ref{branches_g0}. The same will hold at higher genera where the loading function not only has the same monodromies as the tree-level one, but also extra monodromies coming from lack of periodicity along the $A$ and $B$ cycles on the higher-genus surface.

\begin{figure}[t]
	\centering
	\subfloat[A chain $\gamma$ in the twisted chain group.]{\includegraphics[scale=.88]{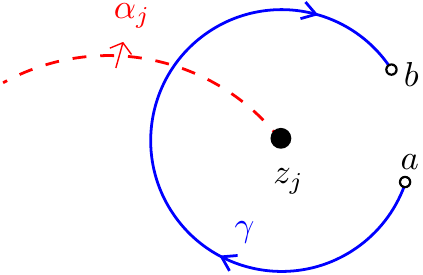}\label{chain}}
	\qquad\qquad
	\subfloat[Cycles in the twisted homology.]{\includegraphics[scale=1]{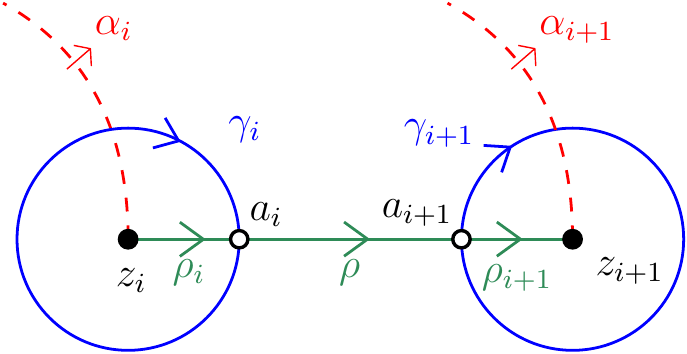}\label{twistedc}}
	\caption{Examples of twisted chains and cycles.}
\end{figure}

The definition of twisted homology groups follows closely that of the usual homology groups. The difference in the twisted case is that the twisted chains keep track of branch cuts that have been crossed. As a concrete example consider the $1$-chain $\gamma$ in Figure~\ref{chain}.
Considered as an element of the first twisted chain group $C_1(\Sigma_{0,n-1},\mathcal{L}_0)$, $\gamma$ carries a section $\mathfrak{s}$ of $\mathcal{L}_0$, therefore, when it crosses the branch cut, this section gets multiplied by the respective monodromy $\mathfrak{s} \mapsto \alpha_j\mathfrak{s}$.
In this picture, $\gamma$ does not live on the universal cover but on the surface $\Sigma_{0,n-1}$. This simplification comes with a trade-off, which is the necessity of keeping track of relative phases along the $1$-chain, i.e., which branch cuts were crossed and how as discussed in the introduction above.

A twisted boundary operator $\bdry: C_1(\Sigma_{0,n-1},\mathcal{L}_0)\rightarrow C_0(\Sigma_{0,n-1},\mathcal{L}_0)$ can be defined such that its action on $\gamma$ reads
\begin{equation}
 \bdry\gamma := \alpha_j b - a
\end{equation}
Note that if $a=b$ this $1$-chain would be a cycle in the usual homology $H_1(\Sigma_{0,n-1},\Z)$, but in the twisted homology $H_1(\Sigma_{0,n-1},{\cal L}_0)$, $\gamma$ is not a cycle unless $\alpha_j=1$. Some cycles in the usual homology are also cycles in the twisted homology, for example, the cycle $\rho_i{+}\rho{+}\rho_{i+1}$ in Figure~\ref{twistedc} is also a cycle in the twisted homology as it does not cross any branch cuts.\footnote{Strictly speaking, since $\rho_i {+} \rho {+} \rho_{i+1}$ is non-compact it is an element of a \emph{locally-finite} twisted homology $H_{1}^{\text{lf}}(\Sigma_{0,n-1},{\cal L}_0)$, but we gloss over this distinction for the purposes of this paper.}
Another interesting twisted cycle is built out of the two circles around adjacent punctures, joined by a line, see Figure~\ref{twistedc},
\begin{equation}
  \label{eq:poch-like} \tilde \gamma := \frac{\gamma_i}{\alpha_i-1}+\gamma-\frac{\gamma_{i+1}}{\alpha_{i+1}-1}.
\end{equation}
Using the fact that $\bdry \gamma_i = (\alpha_i{-}1)a$, and $\bdry \rho = b-a$, it is immediate but instructive to check that its boundary, in the twisted sense, is empty:
\begin{equation}
\bdry\tilde \gamma
  =a+(b-a)-b=0
\end{equation}
and therefore it is a cycle in the twisted homology.
These kinds of cycles are related to the regularization of tree-level string integrals using Pochhammer contours \cite{Mizera:2017cqs}, see also \cite{yoshida2013hypergeometric}, that can be interpreted as implementing Feynman's $i\varepsilon$ prescription in string theory \cite{Witten:2013pra}. They appear naturally in the context of twisted homology, where it is possible to show that $\rho_i{+}\rho{+}\rho_{i+1}$ and $\tilde{\gamma}$ are related by a regularization map in the twisted homology \cite{Mizera:2017cqs}.

\begin{figure}
  \centering
  \includegraphics[scale=0.9]{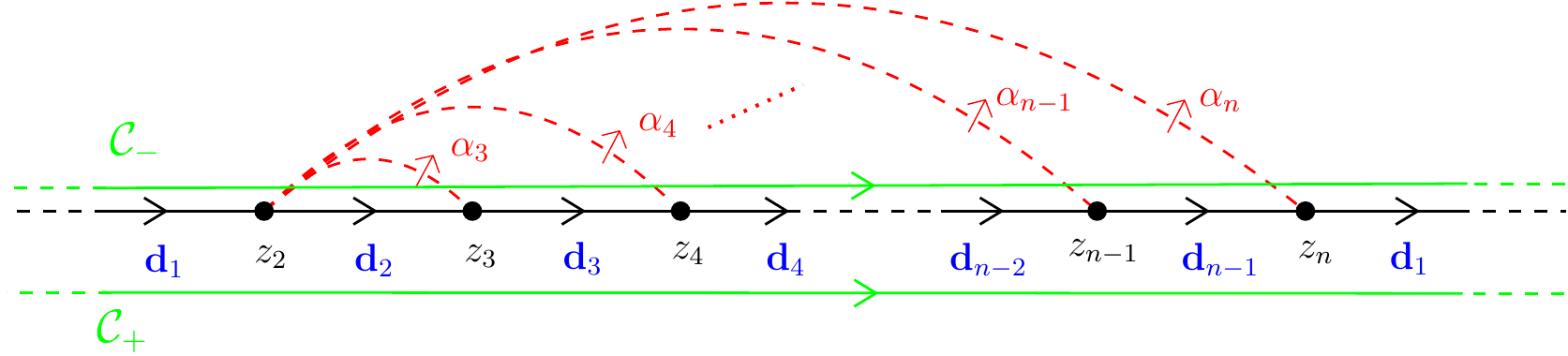}
  \caption{Contours producing boundary relations on the twisted homology. The left- and right-most ends of the horizontal contours are connected, since they are defined on $\CP^1$.}
\label{relations_g0}
\end{figure}

The twisted homology groups $H_d(\Sigma_{0,n-1},\mathcal{L}_0)$ are then defined by the quotient of closed chains by exact chains with respect to the twisted boundary operator.
We will find that this simple change in viewpoint on multi-valued functions; from working on infinite covering space to working on a space with extra structure (the local system), makes transparent and much simpler many properties of the string amplitudes.
In particular, the monodromy relations can now be re-interpreted as boundary relations directly in the twisted homology $H_1(\Sigma_{0,n-1},\mathcal{L}_0)$. In Figure~\ref{relations_g0}, the cycle $\mathbf{d}_1$ denotes the original cycle from which the amplitude~\eqref{tree_open} is analytic continued. The other cycles would represent amplitudes with puncture $z_1$ in different ordering but the branch choice for~\eqref{loading_ex} is only appropriate for the original amplitude where it was analytic continued from. Fortunately, the difference between this and the physical branch (coinciding with \eqref{tree_open}) is a simple phase: \begin{align}\label{phys_branch_g0}
 \int \prod_{i=2}^{n-3}d z_i\int_{\mathbf{d}_i} \!\!dz_1\, T_0(z_1)\,\varphi &=e^{\pi i k_1\cdot\sum_{j=2}^ik_j}\int \prod_{i=2}^{n-3}d z_i\int_{\mathbf{d}_i}dz_1\prod_{j=2}^n|z_j{-}z_i|^{k_i\cdot k_j}\varphi\nn\\
 &=:e^{\pi i k_1\cdot\sum_{j=2}^ik_j}\A(2\cdots i, 1, i{+}1\cdots n).
\end{align}

The relations on the twisted homology are given by the contours $\mathcal{C}_\pm$, which can be written as boundaries, $\mathcal{C}_\pm = \partial \mathcal{D}_\pm$ of two 2-chains $\mathcal{D}_\pm$, and are thus homologous to zero. At the level of the homology, this gives the following two relations:
\begin{subequations}
\begin{align}
  \mathbf{d}_1+\sum_{i=2}^{n-1}\prod_{j=i+1}^{n-1} \alpha_j^{-1}\, \mathbf{d}_i=0\,,\label{eq:Cm}\\
  \sum_{i=1}^{n-1}\mathbf{d}_i=0\,,\label{eq:Cp}
\end{align}
\label{eq:Cpm}
\end{subequations}
for $\mathcal{C}_-$ and $\mathcal{C}_+$, respectively. The coefficients $\alpha_i$ are defined as in eq.~\eqref{eq:alpha-i-def}.

These relations on the twisted homology can be rewritten using \eqref{phys_branch_g0} in terms of relations for the amplitudes. This induces new phases, and the relations read
\begin{align}\label{eq:tree_rel}
 \A(12\cdots n)+e^{\pm\pi i k_1\cdot k_2}\A(213\dots n)&+\cdots
 +e^{\pm\pi i k_1\cdot\sum_{j=2}^i k_j}\A(2,\dots, i,1,i{+}1,\dots, n)\nn\\
 &+\cdots+e^{\pm\pi ik_1\cdot\sum_{j=2}^{n-1}k_j}\A(2,\dots, n{-}1,1,n)=0\,,
\end{align}
for $\mathcal{C}_-$ and $\mathcal{C}_+$, respectively. These bring down the number of independent $\mathbf{d}_i$ contours from $n{-}1$ to $n{-}3$ and were already known in the early literature on dual-models \cite{Plahte:1970wy}. From the modern perspective, they give rise to Kleiss--Kuijf (KK) \cite{Kleiss:1988ne} relations and fundamental BCJ relations ~\cite{Feng:2010my} in the field theory limit.

The relations~\eqref{eq:tree_rel} hold when all points except $z_1$ are held fixed. In principle, there could be extra relations on $\mathcal{M}_{0,n}$ which are not built out of the ones given above, but this turns out not to be the case. The mathematical argument for it comes from computing the dimension of the twisted homology group $H_{n-3}(\mathcal{M}_{0,n},\mathcal{L}_0)$. For generic local coefficients $\alpha_j$ the twisted homology groups $H_d(\mathcal{M}_{0,n},\mathcal{L}_0)$ vanish for $d\neq n{-3}$ \cite{aomoto1975vanishing}.\footnote{The physical intuition is that only on exactly degenerated kinematics (when a propagator goes on-shell, soft or forward limit, etc.) the scattering amplitude factorizes. Accordingly, the local systems becomes degenerate and the dimension of $H_{n-3}({\cal M}_{0,n},{\cal L}_0)$ decreases at a cost of making lower-degree twisted homologies ($d<n{-}3$) non-trivial, in a way that preserves $\chi({\cal M}_{0,n}) = \sum_{d} (-1)^d \dim H_d({\cal M}_{0,n},{\cal L}_0)$.}
Since ${\cal L}_0$ is flat, the alternating sum of dimensions of twisted homology groups is equal to the Euler characteristic of the space
\begin{equation}
\chi({\cal M}_{0,n},{\cal L}_0) = \chi({\cal M}_{0,n})=\sum _{d}(-1)^{d}\mathrm{dim }\,H_d(X,{\mathcal {L}_0})\label{eq:euler-char}
\end{equation}
The Euler characteristic of ${\cal M}_{0,n}$ is given by 
\begin{align}
\chi(\mathcal{M}_{0,n}) &= \prod_{k=3}^{n} \chi(\Sigma_{0,k-1})\nn\\
&= (-1)^{n-3}(n-3)!, 
\end{align}
This follows from the fact that $\chi(\Sigma_{0,k-1}) = 3{-}k$ and that the Euler characteristics on a fibre bundle multiply.
Therefore, given that only the $d=n{-}3$ case is non-vanishing in \eqref{eq:euler-char}, we find
\begin{equation}
\text{dim}\;H_{n-3}(\mathcal{M}_{0,n},\mathcal{L}_0)=(n-3)!.  
\end{equation}
Choosing a basis of $H_{n-3}(\mathcal{M}_{0,n},\mathcal{L}_0)$ such that every element corresponds to a particular choice of $\Delta(\alpha)$ from \eqref{Delta-alpha} fixes the number of independent tree-level open string amplitudes to $(n-3)!$. This argument shows that there are no further (co)homological relations between generic string amplitudes; there could still be other relations between amplitudes of particular string theories, but they would not apply to an \textit{arbitrary} string theory like the monodromy relations do.

A similar result regarding the minimal basis of tree-level string amplitudes was argued for using only the monodromy relations in \cite{BjerrumBohr:2009rd,Stieberger:2009hq}.
The argument there is quite different to ours, as it makes use of only one of the relations in~\eqref{eq:tree_rel} together with its complex conjugate (assuming that partial amplitudes are real), as well as their permutations obtained by relabelings, to arrive at the number $(n{-}3)!$. In order to correctly generalize monodromy relations and basis counting to higher genera it is important to realize that \eqref{eq:tree_rel} contains two separate independent relations.

\subsection{Generalization to higher genera}
\label{sec:Twisted_g}

The first difference at higher genus ($g\geq1$) is that due to zero modes of the string worldsheet action, there are extra contributions to the worldsheet propagator: they give rise to new phases upon analytic continuation. Some of these phases are unwanted; they depend on the surface moduli or the location of other particles, obstructing an interpretation in terms of twisted homology on the configuration space.
One way to remedy this is to introduce an explicit dependence on loop momenta on the string theory integrand. This is a standard trick employed to obtain chirally-split integrands in closed strings~\cite{Verlinde:1987sd,DHoker:1988pdl} and gives a canonical definition of loop momenta in the field theory limit~\cite{Tourkine:2019ukp} for the Feynman graphs generated by closed and open string amplitudes.

\begin{figure}
	\centering
	\includegraphics[scale=1]{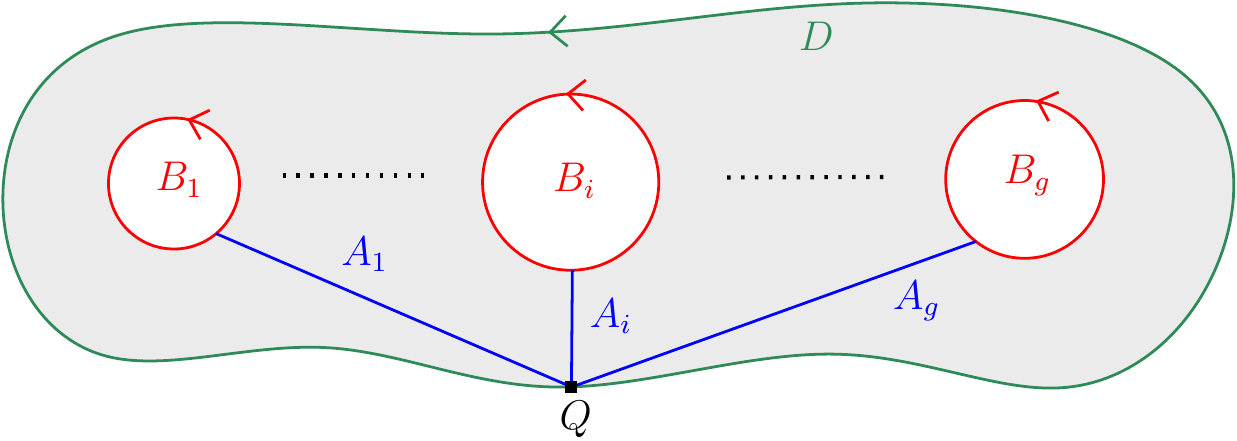}
	\caption{Resulting surface (annulus with g holes) after modding out by the $\mathbb{Z}_2$ involution.}
	\label{fig:open_g_cut}
\end{figure}

In this approach, the integrand of a generic genus-$g$ open-string amplitude has as a universal piece, the Koba--Nielsen factor, supplemented with loop-momentum dependent terms: \begin{equation}\label{KB_g}
 e^{-2 \pi \sum_{I=1}^g\ell_I\cdot \sum_{i=1}^n k_i \Im \int_{P}^{z_i}\omega_I}\prod_{1 \leq i<j \leq n}|E(z_i,z_j)|^{k_i\cdot k_j}\,.
\end{equation}
In this expression, $\ell_I$ are the $g$ loop momenta, $E(z,w)$ is the genus-$g$ prime form (see below and Appendix~\ref{prime_app}), and $P$ is any point except for $Q$ where the $A$-cycles meet, as in Figure~\ref{fig:open_g_cut}. Details on this construction were recently given in \cite{Tourkine:2019ukp}, which are essential to understand the meaning of the loop monodromy relations.
\begin{figure}
	\centering
	\includegraphics[scale=1]{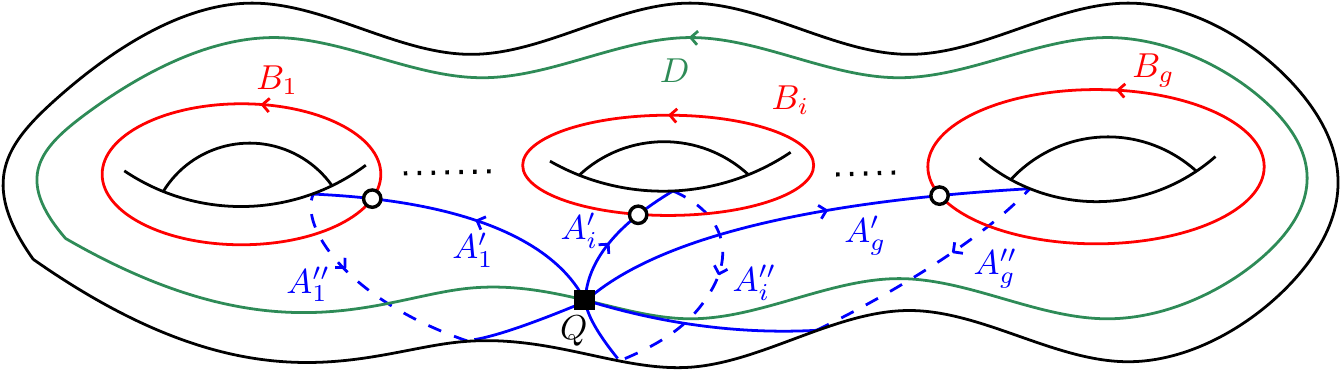}
	\caption{Genus-$g$ surface cut along red and blue cycles.}
	\label{fig:gcut}
\end{figure}

The introduction of the loop momentum requires a choice of representative $A_I$-cycles on the surface along which each loop momentum is measured, and for an open string the points $\{z_i\}$ are inserted along the various boundaries of the worldsheet. In this work we will only be interested in planar, orientable open string worldsheets, since these are conceptually the simplest ones to work with. We do not expect any difficulties in generalizing the framework set up here to more general cases.

When defined as above, i.e., with the $A_I$-cycles touching at the point $Q$, the loop momenta of Feynman integrands obtained from the field theory limit are globally defined throughout all graphs~\cite{Tourkine:2019ukp}, even those of different topologies and non-planar ones. It is important to have an unambiguous definition of the loop momenta since we will see later that higher-genus monodromy relations contain loop momentum-dependent phases. With the choice above these relations will hold identically, that is, without the need of a shift of loop momentum between terms. 

In this representation the string integrand is not the same when approaching the $A_I$-cycles from opposite sides; only after a shift in loop momentum it can be identified as the same. That means that if we want to work at fixed loop momentum, then we must give up working on the original surface, and work instead on a new surface made out of the original one by cutting along the $A_I$-cycles, while not identifying these new boundaries introduced along the cuts. 
On the other hand, if one does not care about working up to shifts in loop momentum then these pieces can be identified giving back the original surface. Both are valid options but in this paper we will mostly work at fixed loop momenta. This choice has implications on what can be considered a minimal basis of integrands. We elaborate on this in the discussion.

\begin{figure}
  \centering
		\includegraphics[scale=1]{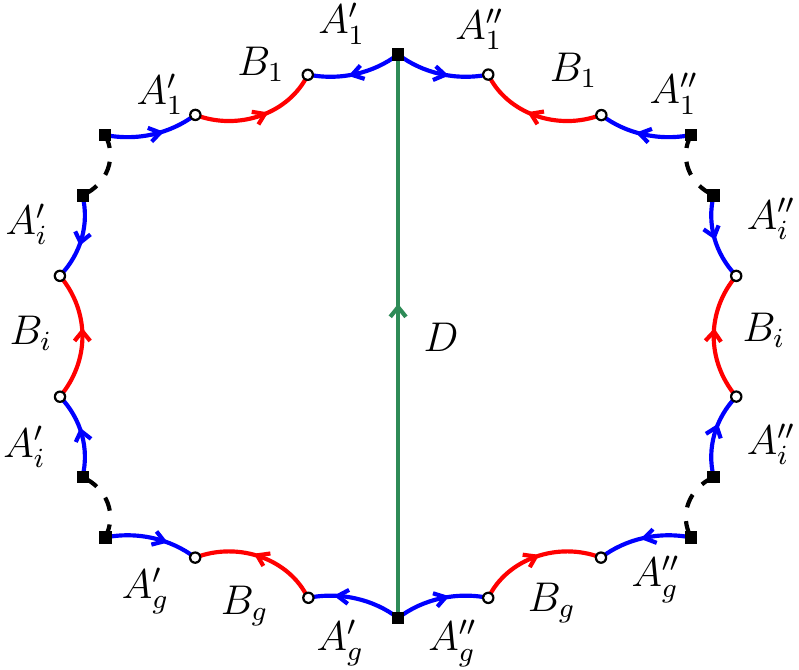}
		\caption{Cut genus $g$ surface.}
                \label{fig:gcut_2}
              \end{figure}

We are interested in analytically continuing the integrand~\eqref{KB_g} in one variable $z_1$, keeping all others fixed. We proceed in analogous fashion to the tree-level case, viewing the open string worldsheet as a ``halved'' closed string worldsheet. That is, for particular values of the modular parameters the open-string worldsheet is a $\mathbb{Z}_2$-projection of the closed-string worldsheet. A way to implement this is to consider cutting a genus-$g$ surface along the cycles $\{A_I\}$ and $\{B_I\}$ depicted on Figure~\ref{fig:gcut}.
The resulting cut surface is symmetric along the cycle $D$, so we can project along it to obtain a surface with an extra boundary along $D$, see Figure~\ref{fig:gcut_2}, and boundaries along the identified $\{B_I\}$ cycles. As discussed above, by allowing shifts of the loop momenta we can identify the integrand along each piece of $A_I$-cycle connected by a $B_I$-cycle and reconstruct the open worldsheet as in Figure~\ref{fig:open_g_cut}.

We can now define on the cut surface the local system associated to the function
\begin{equation}\label{KB_g_loading}
 T_g(z_1) := e^{-2 \pi i\sum_{I=1}^g\ell_I\cdot k_1\int_{P}^{z_1}\omega_I}\prod_{i=2}^{n}E(z_i,z_1)^{k_i\cdot k_1}.
\end{equation}
The monodromies of this function are easy to find; the prime form $E(z_i,z_1)$ behaves locally like $(z_i{-}z_1)$ so the monodromies around the position of the particles are the same as at tree-level, $\alpha_i=e^{-2 \pi i k_i\cdot k_1}$. The new monodromies come from transporting $z_1$ along an $A_I$-cycles giving $\beta_i=e^{-2\pi i k_1\cdot (\ell_i-\ell_{i-1})}$. This is represented in terms of the local system depicted in Figure~\ref{all_genus} later in the text, where crossing the dashed red lines corresponds to multiplication by the associated monodromy, or its inverse depending on the direction dictated by the red arrows.
In the rest of the paper we use this construction to derive monodromy relations for string theory integrands, starting with a detailed genus-one calculation which serves to exemplify the new features that appear when generalizing the genus-zero calculations. There are essentially no new features appearing at genus higher than one; there the calculations are abbreviated.

\paragraph{Remark 1.} In the above we have ignored the actual shape of the surface, that is, its complex structure, and worked with purely topological objects. Taking into account the actual dependence on the surface modular parameters the above procedure is only valid for certain values of these parameters, at least if one wants to recover a physical open string worldsheet. For example, at genus one the open-string possesses a modular parameter which is purely imaginary; to obtain this worldsheet we should start with a closed string whose worldsheet also has purely imaginary modular parameter. Below, we will always be working at fixed values of the modular parameters and we will always assume that they have been fixed to the values allowed by the open string.\footnote{This requirement can be relaxed if an interpretation as string-theory integrands is not required.}

\paragraph{Remark 2.} There are many ways to slice open a genus $g$ surface in order to obtain a simply-connected region of the plane. In particular, the polygon of Figure~\ref{fig:gcut_2}, prior to identifying both sides along the $D$ cycle, is not the standard 4g-sided polygon model for closed surfaces of genus $g$ (see, e.g., \cite{farkas2012riemann}); our polygon has $6g$ edges. One can define a local system for surfaces cut in other ways and calculate the relations in their respective twisted homology. These will be equivalent to the relations we obtain in this paper, up to possible shifts of the loop momenta.

Our choice of cut surface simply ended up being the most convenient for calculations. Similarly, there are several ways to define local systems on the same cut surface, but as long as they encode the monodromy properties of \eqref{KB_g_loading}, their relations in their respective twisted homology are equivalent to the relations we obtain with our local system.

\section{Genus one}
\label{sec:genus_one}

In this section we work out in full detail how to use the set up defined above to find the monodromy relations in the one-loop case. While these are known in the literature~\cite{Tourkine:2016bak,Hohenegger:2017kqy}, there was originally a discrepancy between the relations obtained in each of these works, related to whether or not some integrals along the bulk of the open string graphs cancel in the non-planar amplitudes \textit{after} loop-momentum integration. Hence we take the opportunity to not only reproduce these computations in our set up, but also to present the resolution of this discrepancy.

Let us consider the contribution to open string scattering coming from orientable genus-one Riemann surfaces, i.e., annuli with puncture insertions on both boundaries. We start by embedding it on a torus, $\Sigma_{1} := \C/(\Z {+} \tau \Z)$, with purely imaginary modular parameter $\tau  = i t \in i\R_+$.
We denote positions of the vertex operators with $z_j = x_j + i y_j$, where $y_j\in (0,t]$ and $x_j \in \{0,\frac{1}{2}\}$, depending on the boundary it is inserted. Later we will relax this and consider $x_j\in [-1/2;1/2]$.
We use the $U(1)$ redundancy on the annulus (extended to $U(1){\times}U(1)$ on the torus) to fix the position of one puncture, say $z_m$, to $z_m = \tau$. When all punctures are inserted on a single boundary, the corresponding amplitude is called \emph{planar}, and otherwise it is \emph{non-planar}.

\subsection{Loop integrands}

The Green's function on the annulus depends on whether the punctures $z_j$ and $z_k$ are inserted on the same or different boundary and is given by
\begin{equation}
  \label{eq:G-def}
  G(z_j,z_k) = -G_1(z_j,z_k)-\pi \frac{\Im(z_j-z_k)^2}{\Im \tau}
\end{equation}
where
\begin{equation}
G_1(z_j, z_k) = \begin{dcases}
	\log \left| \frac{\vartheta_1(i y_j {-} i y_k )}{\vartheta_1^\prime(0)} \right| \qquad\text{for}\qquad |x_j {-} x_k| =0,\\		
	\log \left| \frac{\vartheta_2(i y_j {-} i y_k )}{\vartheta_1^\prime(0)} \right| \qquad\text{for}\qquad |x_j {-} x_k| = \frac{1}{2}.
      \end{dcases}
      \label{eq:G1-def}
\end{equation}
Here $\vartheta_1(z)$, $\vartheta_2(z)$ are Jacobi theta functions, related by
\begin{equation} \label{eq:theta1-theta2} \vartheta_1(z+\frac{1}{2})=\vartheta_2(z)
\end{equation}
in the conventions of Appendix~\ref{prime_app}. They also have a dependence on $\tau=it$ usually written $\vartheta_i(z|\tau)$ or $\vartheta_i(z,\tau)$, which we do not write explicitly to make the formulas lighter.
The Green's function on the annulus also possesses a zero-mode term responsible for guaranteeing double-periodicity at one loop, which can be rewritten in terms of a stringy loop momentum. In the closed string sector it makes manifest chiral splitting -- the property that the closed integrand is a holomorphic square of the open string integrand \cite{Verlinde:1986kw,DHoker:1988pdl,DHoker:1989cxq}.

The loop momentum $\ell^\mu\in\mathbb{C}^D,\,\mu=1,\dots,D$ ($D$ is the number of non-compact space-time dimensions), is the average of the string's momentum through a given A-cycle, see \cite{Tourkine:2019ukp} for a detailed definition. In this fashion, an open-string loop integrand is given by
\begin{equation}
\label{loop-integrand-genus-one}
\I(\alpha | \beta) := \int_{0}^{\infty} \!\! dt \int_{\Delta(\alpha | \beta)} \prod_{\substack{j=1\\ j\neq m}}^{n}\! dz_j\, e^{-\pi t \alpha' \ell^2} e^{-2\pi \alpha' \ell \cdot \sum_{j=1}^{n}\! k_j y_j} \!\!\! \prod_{1\leq j < l \leq n}\!\!\!\! e^{\alpha' k_j \cdot k_l\, G_1 (z_j, z_l)} \; \varphi(\ell^\mu, z_j, \tau).
\end{equation}
Scattering amplitudes are obtained by integrating over $\int d^D \ell$ (usually, the cycle in $\mathbb{C}^D$ is chosen to be the Wick-rotated real slice, $\ell^0\to -i\ell^D$ and $\ell^1,\dots,\ell^D\in \mathbb{R}^D$). Here $\alpha$ and $\beta$ denote the orderings of punctures on both boundaries such that $x_j = 0$ for all $j\in \alpha$, and likewise $x_j = \frac{1}{2}$ for all $j \in \beta$. In our gauge fixing we set $\alpha(m)=m$ and $|\alpha|=m$ without loss of generality. The integration domain $\Delta(\alpha|\beta)$ is given by the $(n{-}1)$-dimensional non-compact cycle
\begin{align}
\Delta(\alpha|\beta) := &\{ 0 < y_{\alpha(1)} < y_{\alpha(2)} < \dots < y_{\alpha(m)} {=} t\} \nn\\
                        &\cap \{  t > y_{\beta(1)} > y_{\beta(2)} > \dots > y_{\beta(n-m)} > 0 \}\,,
                          \label{Delta-alpha-beta}
\end{align}
(see also the right-hand side of Figure~\ref{torus_non_planar} below).
The remaining part of the integrand $\varphi$ is a single-valued function encoding the matter content of the vertex operators, whose precise form is irrelevant for our purposes as we are interested in the multi-valuedness of the integrand only.
Given that the integrand \eqref{loop-integrand-genus-one} is multiplied by the translation-invariant integration measure $d^D \ell$, one can check that the scattering amplitudes are invariant under simultaneous shifts $\ell^\mu \to \ell^\mu {-} k_j^\mu$ and $z_j \to z_j {+} \tau$ for any $j$ provided that
\begin{equation}
\varphi(\ell^\mu {-} \textstyle\sum_j n_j k_j^\mu,\, z_j{+}n_j \tau,\, \tau ) = \varphi(\ell^\mu, z_j, \tau), \qquad n_j \in \Z.\label{eq:phi-shift}
\end{equation}
The integrands are also invariant under shifts $z_j \to z_j {+} 1$ provided that $\varphi$ has this property. For a review of open-string integrals at genus zero and chiral splitting, see \cite{Mafra:2018nla}. For clarity, in the remainder of this section we set $\alpha' = 1$. In these unit the masses of external particles are $k_j^2 \in \Z$ for all $j$.

The cycles \eqref{Delta-alpha-beta} are defined on the configuration space $F_{1,n}$ of $n$ points (one of which is always fixed to $0$ or $\tau$ for instance) on a torus with constant modular parameter $\tau$. Consider the \textit{forgetful map} $F_{1,n} \to F_{1,n-1}$, which removes a puncture. Consider further a fibre of this map: it is not hard to see that it is equal to a torus with $n{-}1$ removed points, $\Sigma_{1,n-1} := \Sigma_1 {-} \{z_2,\dots,z_n\}$, which we can parametrize by a coordinate $z_1 = x_1 + i y_1$. Without loss of generality let us fix $m{-}1$ points (for $2 \leq m \leq n$) on the first boundary, and $n{-}m$ on the other, both in the canonical order. The integral from \eqref{loop-integrand-genus-one} can be written as
\begin{equation}\label{I-alpha-beta}
\I(\alpha | \beta) = \int_{0}^{\infty} \!\! dt \int_{\widehat{\Delta}(\alpha | \beta)} \prod_{\substack{j=2\\ j\neq m}}^{n}\! dz_j\, e^{-\pi t \ell^2} e^{-2\pi \ell \cdot \sum_{j=2}^{n} k_j y_j} \!\!\! \prod_{2\leq j < k \leq n}\!\!\!\! e^{k_j \cdot k_k\, G_1 (z_j, z_k)}\; \widehat{\I}(\alpha|\beta),
\end{equation}
where $\widehat{\Delta}(\alpha|\beta)$ is projection under the forgetful map. To write down $\widehat{\I}(\alpha|\beta)$ we distinguish two cases depending on where $z_1$ is integrated over. For $x_1 = 0$ we have
\begin{align}
&\widehat{\I}(2,3,\dots,i,1,i{+}1,\dots,m|m{+}1,\dots,n)\nn\\
&= i \int_{y_i}^{y_{i+1}} dy_1\, e^{-2\pi k_1 \cdot \ell\, y_1} \prod_{j=2}^{m} \left| \vartheta_1(i y_j {-} iy_1) \right|^{k_1 \cdot k_j} \!\!\! \prod_{j=m+1}^{n}\!\!\! \left| \vartheta_2(iy_j {-} iy_1) \right|^{k_1\cdot k_j} \varphi(\ell^\mu, z_j, \tau),
\end{align}
where $2 \leq i{+}1 \leq m$ and in the edge case $i{+}1=2$ the integral is over the interval $(0,y_2)$. Similarly, for $x_1 = \frac{1}{2}$:
\begin{align}
&\widehat{\I}(2,\dots, m|m{+}1,\dots, i,1,i{+}1,\dots, n)\nn\\
&= i \int_{y_i}^{y_{i+1}} dy_1\, e^{-2\pi k_1 \cdot \ell\, y_1} \prod_{j=2}^{m} \left| \vartheta_2(i y_j {-} iy_1) \right|^{k_1 \cdot k_j} \!\!\! \prod_{j=m+1}^{n}\!\!\! \left| \vartheta_1(iy_j {-} iy_1) \right|^{k_1\cdot k_j} \varphi(\ell^\mu, z_j, \tau)
\end{align}
for $m \leq i \leq n$. In the edge case $i=n$ the integration domain is $(y_n,0)$. The only difference between the two expressions is the switch $\vartheta_1 \leftrightarrow \vartheta_2$ and the integration domain. Note that we dropped the normalization $\vartheta_1'(0)$ from the Green's functions by momentum conservation and integer mass condition.

When linear relations between $\widehat{\I}(\alpha|\beta)$'s are independent of the positions of the remaining punctures $z_{j\neq 1}$ and $\tau$, as will be the case here, they carry over to relations between the full loop integrands $\I(\alpha|\beta)$. Should the coefficients of such relations be further independent of $\ell^\mu$, they also extend to scattering amplitudes. This will be the case in particular for the generalization of the Bern--Dixon--Dunbar--Kosower (BDDK) relations~\cite{Bern:1994zx}, which are relations that hold at the field theory level, for which all the coefficients reduce to $\pm1$. 

\subsection{Local system and monodromy relations}

\begin{figure}[!ht]
	\begin{center}
		\includegraphics[scale=1]{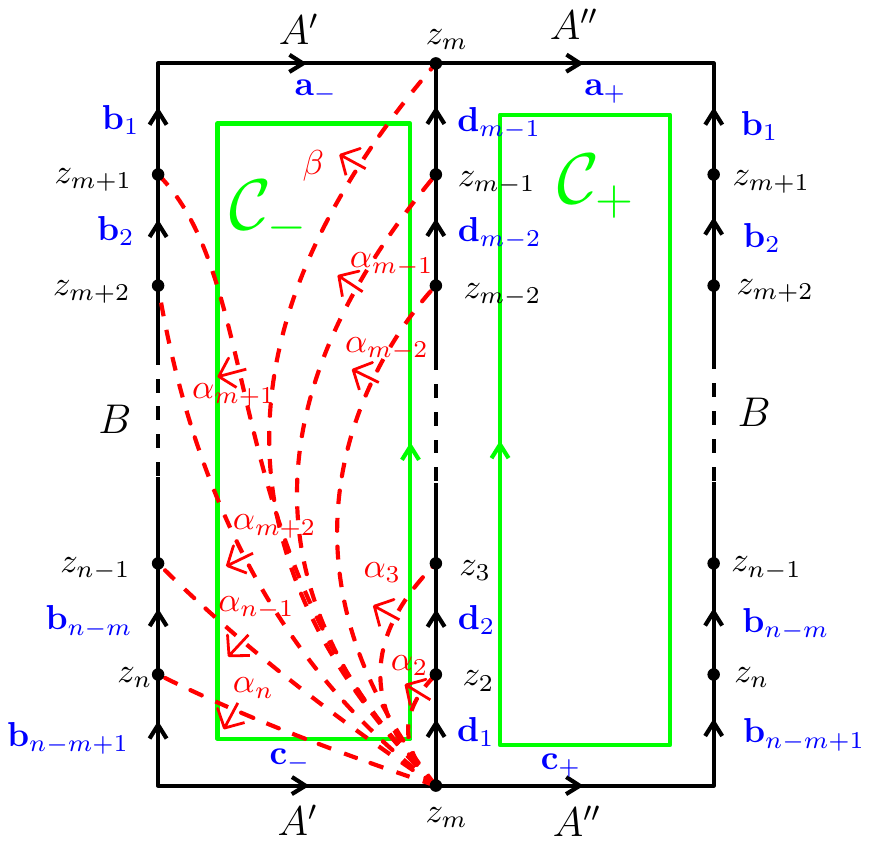}
		\caption{\label{torus_non_planar}Integration chains $\mathbf{a}_{\pm}, \mathbf{b}_j, \mathbf{c}_{\pm}, \mathbf{d}_j$ and the local system $\mathcal{L}_1$ defined by the phases $\alpha_j$ and $\beta$ on a cylinder. Identifying $\mathbf{a}_\pm = \mathbf{c}_\pm$ gives a torus.}
	\end{center}
\end{figure}

In order to perform contour manipulations on the loop integrands we need to first express them as integrals over holomorphic functions in $z_1$, which comes at a price of introducing further multi-valuedness. We start by defining
\begin{equation}\label{T1}
T_1(z_1) := e^{2\pi i k_1\cdot \ell\, z_1} \prod_{j=2}^m \left(-i \vartheta_1(iy_j{-}z_1)\right)^{k_1\cdot k_j}\prod_{j=m+1}^n\vartheta_2(iy_j{-}z_1)^{k_1\cdot k_j}.
\end{equation}
It is designed so that the integral of $T_1(z_1) \varphi$ on the interval $(0,y_2)$ agrees with $\widehat{\I}(12\dots m|\allowbreak m{+1}\dots n)$, including the phase.\footnote{The factor of $-i$ is needed since $\vartheta_1(z) = i |\vartheta_1(z)|$ for $z\in [0,it]$.} Monodromies of the above functions define the local system $\mathcal{L}_1$ on a torus with $n{-}1$ punctures removed. The local system is defined by a choice of branch cuts of \eqref{T1} and associated monodromies labeled $\alpha_j$ and $\beta$, both shown in Figure~\ref{torus_non_planar}.
Using monodromy properties of theta functions we can easily identify:
\begin{equation}\label{local-system-genus-one}
\alpha_j=e^{-2\pi i k_1\cdot k_j},\qquad \beta=e^{-2\pi i k_1\cdot \ell} e^{-2\pi i k_1\cdot k_m}.
\end{equation}
The two twisted cycles $\mathcal{C}_\pm$ in Figure~\ref{torus_non_planar} are contractible to a point and thus zero in the twisted homology $H_1(\Sigma_{1,n-1}, \mathcal{L}_1)$. From them we obtain two monodromy relations:
\begin{align}\label{non_planar_1}
\sum_{i=1}^{m-1}\prod_{j=2}^i\alpha_j\, \mathbf{d}_i &-\beta\prod_{j=2}^{m-1}\alpha_j\, (\mathbf{a}_- +\mathbf{b}_1)\nn\\
&-\beta\prod_{j=2}^{m-1}\alpha_j\sum_{i=2}^{n-m}\prod_{k=m+1}^{m+i-1}\alpha_k\, \mathbf{b}_i-
\beta\prod_{\substack{j=2\\ j\neq m}}^{n}\alpha_j\, (\mathbf{b}_{n-m+1}-\mathbf{c}_-)=0
\end{align}
coming from $\mathcal{C}_- = 0$, and similarly
\begin{equation}\label{non_planar_2}
\sum_{i=1}^{m-1}\mathbf{d}_i+\mathbf{a}_+ - \!\!\!\sum_{i=1}^{n-m+1}\mathbf{b}_i-\mathbf{c}_+ =0
\end{equation}
from $\mathcal{C}_+ = 0$. Translating these two identities into loop integrands is a matter of relating the choice of branch of $T_1(z_1)$ to that of $\widehat{\I}(\alpha|\beta)$ given by the string amplitude. We have
\begin{equation}
\int_{\mathbf{d}_i} \!\!dz_1\, T_{1}(z_1)\, \varphi = e^{\pi i k_1 \cdot \sum_{j=2}^{i} k_j}\, \widehat{\I}(2,3\dots, i,1,i{+}1,\dots,m|m{+}1,\dots,n)
\end{equation}
for $1 \leq i \leq m{-}1$, as well as
\begin{equation}
\int_{\mathbf{b}_{i-m+1}} \!\!\!\!\!\!\!\!\!\!\!dz_1 \, T_{1}(z_1)\, \varphi = - e^{\pi i k_1 \cdot \ell} e^{\pi i k_1 \cdot \sum_{j=2}^{i} k_j} \,\widehat{\I}(2,\dots, m | m{+}1,\dots,i,1,i{+}1,\dots,n)
\end{equation}
for $m\leq i \leq n$.

In addition we have the chains $\mathbf{a}_\pm$ and $\mathbf{c}_\pm$ that enter the relations \eqref{non_planar_1} and \eqref{non_planar_2} which do not have an immediate string integrand interpretation. In order to be precise let us define integrals over these chains by
\begin{equation}
\label{a_chain}
\int_{\mathbf{a}_\pm} \!\! dz_1\, T_{1}(z_1)\, \varphi =: e^{\pi i k_1 \cdot \ell} e^{\pi i k_1 \cdot \sum_{j=2}^{m}k_j}\, \widehat{\J}_{\mathbf{a}_\pm} (2,\dots, m| 1,m{+}1,\dots, n)\,,
\end{equation}
\begin{equation}
\label{c_chain}
\int_{\mathbf{c}_\pm} \!\! dz_1\, T_{1}(z_1)\, \varphi =: e^{\pi i k_1 \cdot \ell}\, \widehat{\J}_{\mathbf{c}_\pm} (2,\dots, m| m{+}1,\dots, n,1)\,.
\end{equation}
The overall normalization is chosen so that the phases of the integrand of each $\widehat{\J}_{\mathbf{a}_\pm/\mathbf{c}_\pm}(\alpha | \beta)$ agree with those of $\widehat{\I}(\alpha|\beta)$ near $z_1 = \frac{1}{2}$ (for $\mathbf{c}_\pm$) and $z_1 = \tau + \frac{1}{2}$ (for $\mathbf{a}_\pm$). This also allows for their unambiguous definition with other choices of branch cuts.\footnote{This choice is motivated by the fact that these combine with the $\mathbf{b}$ chains at the edge of Figure~\ref{torus_non_planar} to form cycles in the twisted homology.} The full integrals $\J(\alpha|\beta)$ are defined in the same way as in \eqref{I-alpha-beta}. The relevance of $\J(\alpha|\beta)$ contributions in monodromy relations was first highlighted in \cite{Hohenegger:2017kqy}.

Putting everything together, \eqref{non_planar_1} and \eqref{non_planar_2} yield the following loop-integrand relations:
\begin{align}\label{string_non_planar}
&\sum_{i=1}^{m-1}e^{\pm\pi i k_1\cdot \sum_{j=2}^{i} k_{j}} \,\I(2,3,\dots, i,1,i{+}1,\dots, m | m{+}1,\dots, n)\nn\\
&\qquad\qquad +\sum_{i=m}^n e^{\pm\pi i k_1\cdot(\ell+ \sum_{j=2}^{i}k_{j})} \,\I(2,\dots, m|m{+}1, \dots, i, 1, i{+}1,\dots, n)\\
&= \mp e^{\pm\pi i k_1 \cdot \ell} \left( e^{\pm\pi i k_1 \cdot \sum_{j=2}^{m} k_j}\J_{\mathbf{a}_\pm}(2,\dots, m|1,m{+}1,\dots, n) - \J_{\mathbf{c}_\pm}(2,\dots, m| m{+}1, \dots, n,1) \right)\nn
\end{align}
for $-$ and $+$ respectively. It is understood that for $i=1$ the first term is $\mathcal{I}(1,2,\dots,m|m+1,\dots,n)$, and similarly for the second term at $i=n$ we have $\mathcal{I}(2,\dots,m|m+1,\dots,n,1)$. The fact that the coefficients do not depend on $z_{j\neq 1}$ and $\tau$ allowed us to remove the hats. Notice that the coefficients in the two relations are, like at tree-level, complex conjugates to each other. This does not necessarily implies that they carry the same information, as in general the loop integrands $\I(\alpha|\beta)$ and the $\J(\alpha|\beta)$ contributions can be complex-valued.

We have verified our relations numerically at finite values of $\alpha'$ by sampling random points in the kinematic and moduli space up to $n=9$, giving us confidence that they are correct. The numerical check consisted of integrating the variable $z_1$ at fixed values of the loop momentum and modulus of the annulus, and fixed punctures $z_j$ for $j\geq2$.

To the best of our understanding, they match those of~\cite{Tourkine:2016bak,Hohenegger:2017kqy} 
with some caveats. In a first version of~\cite{Tourkine:2016bak}, the analytically continued integrand was assumed to be doubly periodic after a shift in loop momentum, but that is not the case. Only in the specific case of the planar integrand this is true due to momentum conservation. In the generic situation, where particles are inserted on both boundaries there are signs that appear when translating the theta functions leading to an overall phase, see Appendix~\ref{prime_app} for details. This phase is what prevents the two bulk cycles from cancelling each other after a shift in loop momentum.
The relations found in~\cite{Hohenegger:2017kqy} contain terms similar to the contributions from \eqref{a_chain} and \eqref{c_chain}, which, to the best of our understanding, match ours.

Using the periodicity properties of the theta functions, it is easy to check that this phase produces an order $\alpha'$ contribution (a constant), which was observed in \cite{Hohenegger:2017kqy}. It is yet unclear why the computation of the Appendix of \cite{Ochirov:2017jby} seems to have missed this constant.

\subsection{Shifts of the loop momentum}
\label{subsec:g_1_l_shift}

As explained at the beginning of this section, amplitude contributions $\int \!d^D\ell\, \I(\alpha|\beta)$ are invariant under simultaneous shifts $\ell^\mu \to \ell^\mu {-} k_1^\mu$ and $z_1 \to z_1 {+} \tau$. Since phases of $\J(\alpha|\beta)$ are fixed to the corresponding $\I(\alpha|\beta)$, this implies
\begin{equation}\label{J-a-c-shift}
\J_{\mathbf{a}_\pm}(2,\dots,m|1,m{+}1\dots,n) \;\cong\; \J_{\mathbf{c}_\pm}(2,\dots,m|m{+}1,\dots,n,1),
\end{equation}
where we introduced the notation $\cong$ to denote equality after shifts of loop momenta. Thus the right-hand side of \eqref{string_non_planar} cancels out only in the case $m{=}n$ (by momentum conservation and $k_1^2 \in \Z$). In this case, the relations \eqref{string_non_planar} become 
\begin{align}\label{string_planar}
&\sum_{i=1}^{n-1}e^{\pm\pi i k_1\cdot \sum_{j=2}^{i} k_{j}} \,\I(2,3,\dots, i,1,i{+}1,\dots, n | \cdot)+e^{\pm\pi i k_1\cdot \ell} \,\I(2,\dots, n|1) \cong 0,
\end{align}
relating $n{-}1$ planar integrands to a single non-planar one. Those are the original relations of \cite{Tourkine:2016bak}, rederived in \cite{Hohenegger:2017kqy}, which, in the field theory limit, produce those of \cite{Boels:2011tp}.

In the absence of cancellation on the right-hand side of \eqref{string_non_planar}, let us consider pairing up the corner contributions coming from the chains $\mathbf{b}_1$ together with $\mathbf{a}_\pm$ into a single object, namely
\begin{equation}
\K_\pm (\alpha | \beta) := \I (\alpha | \beta ) \pm \J_{\mathbf{a}_\pm} (\alpha| \beta)\label{eq:Ka-def}
\end{equation}
for $(\alpha|\beta) = (2,3,\dots, m| 1,m{+}1,\dots, n )$. Similarly, pairing up $\mathbf{b}_{n-m+1}$ with $\mathbf{c}_\pm$ we define
\begin{equation}
\K_\pm (\alpha | \beta) := \I (\alpha | \beta ) \mp \J_{\mathbf{c}_\pm} (\alpha| \beta)\label{eq:Kc-def}
\end{equation}
for $(\alpha|\beta) = (2,3,\dots, m| m{+}1,\dots, n,1 )$. One motivation for the above definitions is that only the combinations $\mathbf{b}_1 \mp \mathbf{a}_\pm$ and $\mathbf{b}_{n-m+1}\pm\mathbf{c}_\pm$ are twisted cycles in $H_1(\Sigma_{1,n-1},\mathcal{L}_1)$, since the individual chains $\mathbf{a}_\pm, \mathbf{b}_1, \mathbf{b}_{n-m-1}, \mathbf{c}_\pm$ have boundaries at either $z_1 = \frac{1}{2}$ or $z_1 = \tau + \frac{1}{2}$.
This lead us to introducing the following notation for these cycles:
\begin{equation}
  \label{eq:k-ac-pm}
  {\bf k}_{{\bf a},\pm} := {\mathbf{b}}_1 \mp \mathbf{a}_\pm,\qquad {\bf k}_{{\bf c},\pm} := {\mathbf{b}}_{n-m+1}\pm\mathbf{c}_\pm.
\end{equation}
Applying these definitions to \eqref{string_non_planar} we obtain two $n$-term identities:
\begin{align}\label{monodromy-with-cycles}
&\sum_{i=1}^{m-1}e^{\pm\pi i k_1\cdot \sum_{j=2}^{i} k_{j}} \,\I(2,3,\dots, i,1,i{+}1,\dots, m | m{+}1,\dots, n)\nn \\
&\qquad\qquad +\sum_{i=m+1}^{n-1} e^{\pm\pi i k_1\cdot(\ell+ \sum_{j=2}^{i}k_{j})} \,\I(2,\dots, m|m{+}1, \dots, i, 1, i{+}1,\dots, n)\\
&+ e^{\pm \pi i k_1 \cdot (\ell + \sum_{j=2}^{m} k_j)} \K_\pm (2,\dots,m | 1,m{+}1,\dots,n) + e^{\pm \pi i k_1 \cdot \ell} \K_{\pm}(2,\dots,m|m{+}1,\dots,n,1) = 0.\nn
\end{align}
This is the most natural form of these identities in order to discuss a minimal basis of loop-level string integrands. That is because the counting, as done at tree-level in the end of section~\ref{sec:review}, counts the number of independent twisted cycles which are not in principle in one-to-one correspondence with string amplitudes.
While this is true at tree-level we have seen that this is not the case at genus one, the $\K$ cycles contain pieces running along the $A$-cycle. We shall comment more on this in the discussion section.

\section{Arbitrary genus}
\label{sec:higher_g}

\subsection{Geometric set-up}
\label{sec:set-up}

The particles on a planar open string $n$-point, $g$-loop amplitude on the annulus with $g$ holes can be distributed along any of its $g+1$ boundaries, which we denote by $D$, $B_1,B_2,\dots,B_g$, with $r_0$, $r_1,r_2,\dots,r_g$ particles, respectively.
These amplitudes are written as $A(\{D\}|\{B_1\}|\dots|\{B_g\})$ where $\{\cdot\}$ is a shorthand for some permutations of the particles on the respective boundary. We pick the integrand of the amplitude $A(1,2\dots r_0|r_0+1\cdots \tilde{r}_1|\cdots|\tilde{r}_{g-1}+1,\cdots \tilde{r_g})$, where $\tilde{r}_i=\sum_{j=0}^ir_j$, to analytic continue and work on the fibre $\Sigma_{g,n-1}$ in the configuration space, where all but one puncture, $z_1$, are held fixed. In this case, the relevant part of the string integrand is
\begin{equation}\label{g2_int}
 e^{-2 \pi i\sum_{I=1}^2\ell_I\cdot k_1\text{Im}\int_{P}^{z_1}\omega_I} \prod_{i=2}^{n}|E_g(z_i,z_1)|^{k_i\cdot k_1},
\end{equation}
where $E_g(z,w)$ is the genus-$g$ prime form
whose properties we review in Appendix~\ref{prime_app}, and $\omega_I$ are the $g$ linearly independent Abelian differentials canonically normalized so that
\begin{equation}
  \label{eq:abelian-diff}
  \oint_{A_I}\omega_J = \delta_{IJ},\qquad \oint_{B_I}\omega_J = \Omega_{IJ},
\end{equation}
where $\Omega_{IJ}$ is the period matrix of the surface, which encodes its complex structure and which we keep fixed in this work.

\begin{figure}
	\begin{center}
		\includegraphics[width=\textwidth]{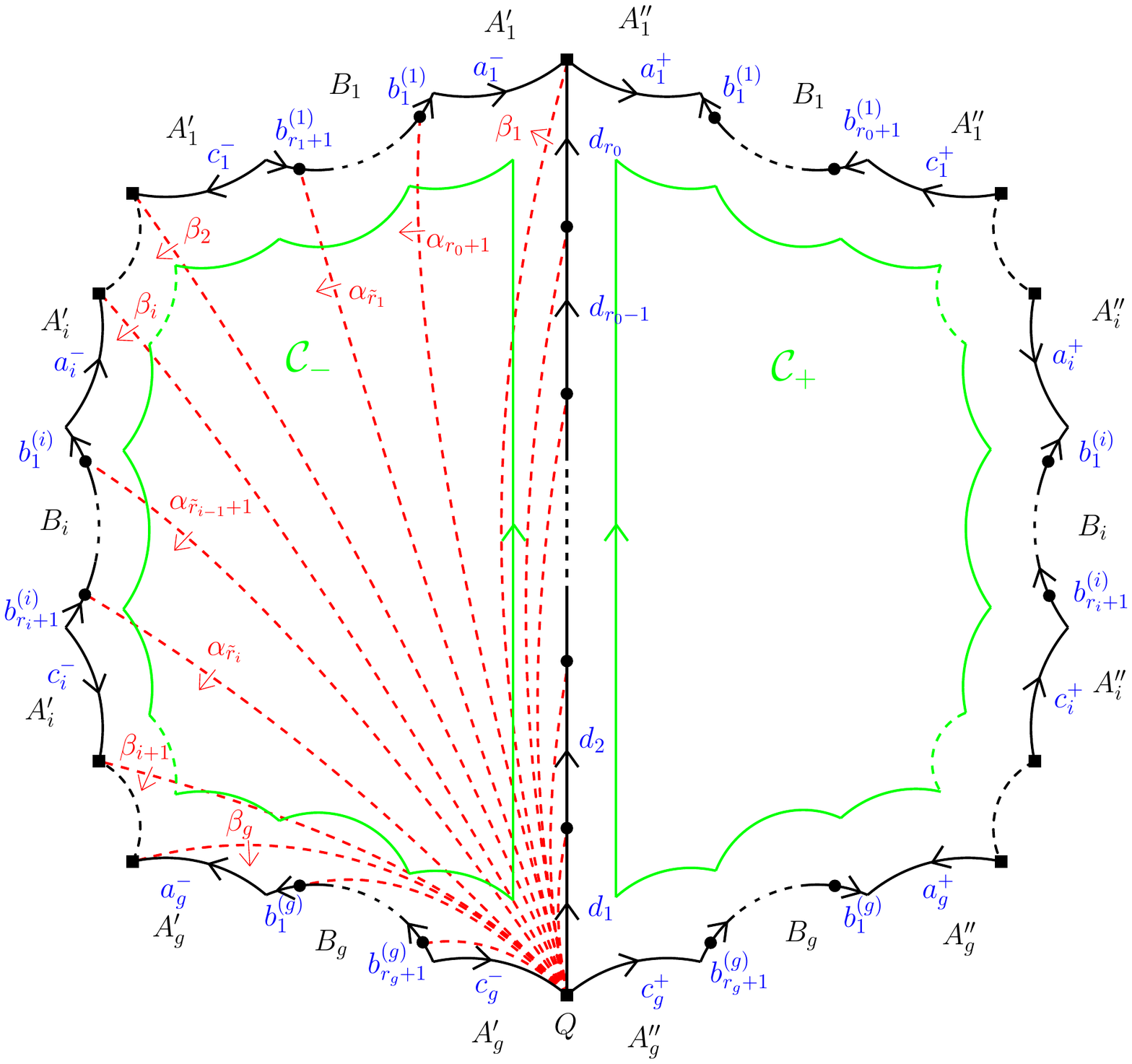}
		\caption{Cycles and local system for the generic non-planar relations on a genus-$g$ surface.}\label{all_genus}
	\end{center}
\end{figure}

The open surface above can be obtained from a closed, genus $g$ surface by cutting it along representative $A$ and $B$-cycles as ain Figure~\ref{fig:gcut} to obtain a polygon on the plane, symmetric along reflections through the cycle $D$ shown on Figure~\ref{all_genus}.
We use this cutting procedure to induce an ordering for the boundaries $D<B_1<\dots< B_g$ which is used to define the analogous of physical branches for the higher genus loading function.
With some abuse of notation write the prime form as $E(iy_i-iy_j)$ when both coordinates are on the same boundary and $y_i>y_j$ (The coordinates $y$ represent a choice of real coordinate along the boundary). This choice corresponds to the physical branch just as defined at genus 1. When particles are on different boundaries the physical branches is as follow:  if $iy_i\in D$ and $iy_j\in B_I$ then write $E(iy_j-iy_i+A_I/2)$ where $A_I/2$ represents the shift due to these particles being on different boundaries, it is the analogous to the shift by $1/2$ at genus 1; if $iy_i\in B_I$ and $iy_j\in B_J$ we take $E(iy_j-iy_i+A_J/2-A_I/2)$ for $I<J$. 

With this notation the loading function is taken as the integrand of $\I(1,2,\dots, r_0|\dots|\{B_g\})$ analytic continued on $z_1$:
\begin{multline}\label{KB_g_detailed}
 T_g(z_1) := e^{2 \pi i\sum_{I=1}^g\ell_I\cdot k_1\int_{P}^{z_1}\omega_I}\prod_{i=2}^{r_0}E(iy_i-z_1)^{k_1\cdot k_i}\times \\ \times\prod_{j=r_0+1}^{r_1}E(iy_j-z_1+A_1/2)^{k_1\cdot k_j}\;\cdots \prod_{p=\tilde{r}_{g-1}}^{\tilde{r_g}}E(iy_p-z_1+A_p/2)^{k_1\cdot k_p}.
\end{multline}
The local system corresponding to this loading function is depicted in Figure~\ref{all_genus} as the dotted red lines, representing the monodromies of the loading function; while the contours appearing in the monodromy relations are in black. Unlike genus zero and one, surfaces of genus $g\geq2$ do not have any continuous automorphisms that leave the complex structure invariant, so no particle needs to be fixed.
At genus one we choose all the red dashed lines of the local system to end on the fixed puncture, but here there is no such fixed puncture. In principle these lines could end anywhere. Nevertheless, there is a choice of point that simplifies the computations, denoted by $Q$, and shown as a square in Figures~\ref{fig:gcut} and~\ref{all_genus}, where the $A$-cycles meet the $D$-cycle. For convenience we choose all our local system cycles to connect to this point.
Other choices lead to different-looking computations, but ultimately to the same monodromy relations once rewritten in terms of the physical branches.

\subsection{Higher-genus monodromy relations}
The relevant relations in the twisted homology are obtained from the homologically trivial contours $\mathcal{C}_\pm$ in Figure~\ref{all_genus}. The relations are
\begin{equation}\label{g_npr_+}
\sum_{i=1}^{m-1}\mathbf{d}_i+\sum_{i=1}^g(\mathbf{a}_i^+-\mathbf{c}_i^+)+\sum_{j=1}^{g}\sum_{i=1}^{r_j+1}\mathbf{b}_i^{(j)}=0
\end{equation}
and the more involved
\begin{multline}\label{g_npr_-}
\mathbf{d}_1 + \sum_{i=2}^{r_0}\prod_{j=2}^{i}\alpha_j \mathbf{d}_i - \sum_{j=1}^g\prod_{q=1}^j\beta_q\tilde{\alpha}_{q-1}(\mathbf{a}_i^-+\mathbf{b}_1^{(j)})\\-\sum_{i=1}^g\prod_{q=1}^i\beta_q\tilde{\alpha}_{q-1}\sum_{j=2}^{r_i}\prod_{l=\tilde{r}_i+1}^{j+\tilde{r}_i-1}\alpha_l \mathbf{b}_j^{(i)}-\sum_{i=1}^g\prod_{q=1}^i\beta_q\tilde{\alpha}_q\tilde{\alpha}_{0}(\mathbf{b}_{r_i+1}^{(i)}-\mathbf{c}_i^-)=0,
\end{multline}
where
\begin{equation}
\tilde{\alpha}_i=\begin{cases}
\prod_{j=2}^{r_0}\alpha_j &\mbox{if }i=0,\\ 
\prod_{j\in \{B_i\}}^{r_i}\alpha_j&\mbox{if }i>0.
\end{cases}
\end{equation}
The local system coefficients are given by the monodromies of~\eqref{KB_g_detailed}
\begin{equation}\label{local-system-genus-g}
\alpha_j=e^{-2\pi i k_1\cdot k_j},\qquad \beta_i=e^{-2\pi i k_1\cdot( \ell_i-\ell_{i-1})}
\end{equation}
with $\ell_0=0$. Using the notation
\begin{equation}
 \int_{{\mathbf a}_i^{\pm}}\!\! dz_1\,T_2(z_1)=e^{\pi i k_1\cdot \ell_i} e^{\pi i k_1\cdot \sum_{j=2}^{\tilde{r}_{i-1}} k_j}\hat{\mathcal{J}}_{\mathbf{a}_i^\pm}(\cdots|1, \tilde{r}_{i-1}+1\cdots \tilde{r}_i|\cdots)
\end{equation}
\begin{equation}
 \int_{\mathbf{c}_i^{\pm}}\!\! dz_1\,T_2(z_1)=e^{\pi i k_1\cdot \ell_i} e^{\pi i k_1\cdot \sum_{j=2}^{\tilde{r}_{i}} k_j}\hat{\mathcal{J}}_{\mathbf{c}_i^\pm}(\cdots| \tilde{r}_{i-1}+1\cdots \tilde{r}_i,1|\cdots)
\end{equation}
the relations \eqref{g_npr_+} and \eqref{g_npr_-} can be rewritten as relations for the integrand in the physical branch as
\begin{multline}
  \hat\I(12\cdots r_0|\cdot)+\sum_{i=2}^{r_0}e^{\pm \pi i k_1\cdot \sum_{j=2}^{i} k_j}\hat\I(2\dots i,1,i{+}1\dots r_0|\cdot)\\
  +\sum_{j=1}^g e^{\pm \pi i k_1\cdot \ell_j}e^{\pm i\pi k_1\cdot \sum_{p=2}^{\tilde{r}_{j-1}}k_p}\hat\I(\cdot|1,\tilde{r}_{j-1}{+}1,\cdots,\tilde{r}_j|\cdot)\\+\sum_{j=1}^g e^{\pm \pi i k_1\cdot \ell_j}e^{\pm i\pi k_1\cdot \sum_{\tilde{r}_{p=2}}^{\tilde{r}_{j-1}}k_p}\sum_{i=\tilde{r}_{j-1}}^{\tilde{r}_j}e^{\pi ik_1\cdot\sum_{p=\tilde{r}_{j-1}}^{i}k_p} \hat\I(\cdot|\tilde{r}_{j-1}+1,\dots i,1,i+1\dots \tilde{r}_j|\cdot)\\
  = \mp\sum_{j=1}^g e^{\pm\pi i k_1\cdot \ell_j} e^{\pm\pi i k_1\cdot \sum_{p=2}^{\tilde{r}_{j-1}} k_p}\bigg(\hat{\mathcal{J}}_{\mathbf{a}_i^\pm}(\cdot|1, \tilde{r}_{j-1}{+}1\cdots \tilde{r}_j|\cdot)\\
  - e^{\pm\pi i k_1\cdot \sum_{p=\tilde{r}_{j-1}}^{\tilde{r}_{j}}\!\!\! k_p}\hat{\mathcal{J}}_{\mathbf{c}_i^\pm}(\cdot| \tilde{r}_{j-1}{+}1\cdots \tilde{r}_j,1|\cdot)\bigg),
  \label{eq:g_amp_rel}
\end{multline}
where, to keep the notation manageable, we only show explicitly the particles on the boundary where particle $1$ is moving.

The relations~\eqref{eq:g_amp_rel} agree with the original ones in~\cite{Tourkine:2016bak} in the planar sector but differ in the non-planar sector in two ways: Firstly, the phases have an accumulative behavior; every time the particle $1$ winds along a new boundary, it keeps the phases from particles inserted on previous boundaries (therefore the planar relations are not sensitive to it by momentum conservation). Secondly, similarly to the one-loop case, the non-planar relations should possess boundary contributions along $A$ cycles with different phases which prevent their cancellation by loop momentum integration. These phases were missed in the original formulae of \cite{Tourkine:2016bak}, leading to the incorrect conclusion that bulk terms could always be removed by loop integration. Contributions from contours along the $A$-cycles were already seen to be present at higher genus in~\cite{Hohenegger:2017kqy}, where it was noted that because of phases differences these do not cancel after integrating out the loop momenta.

\begin{figure}[ht]
	\centering
	\includegraphics{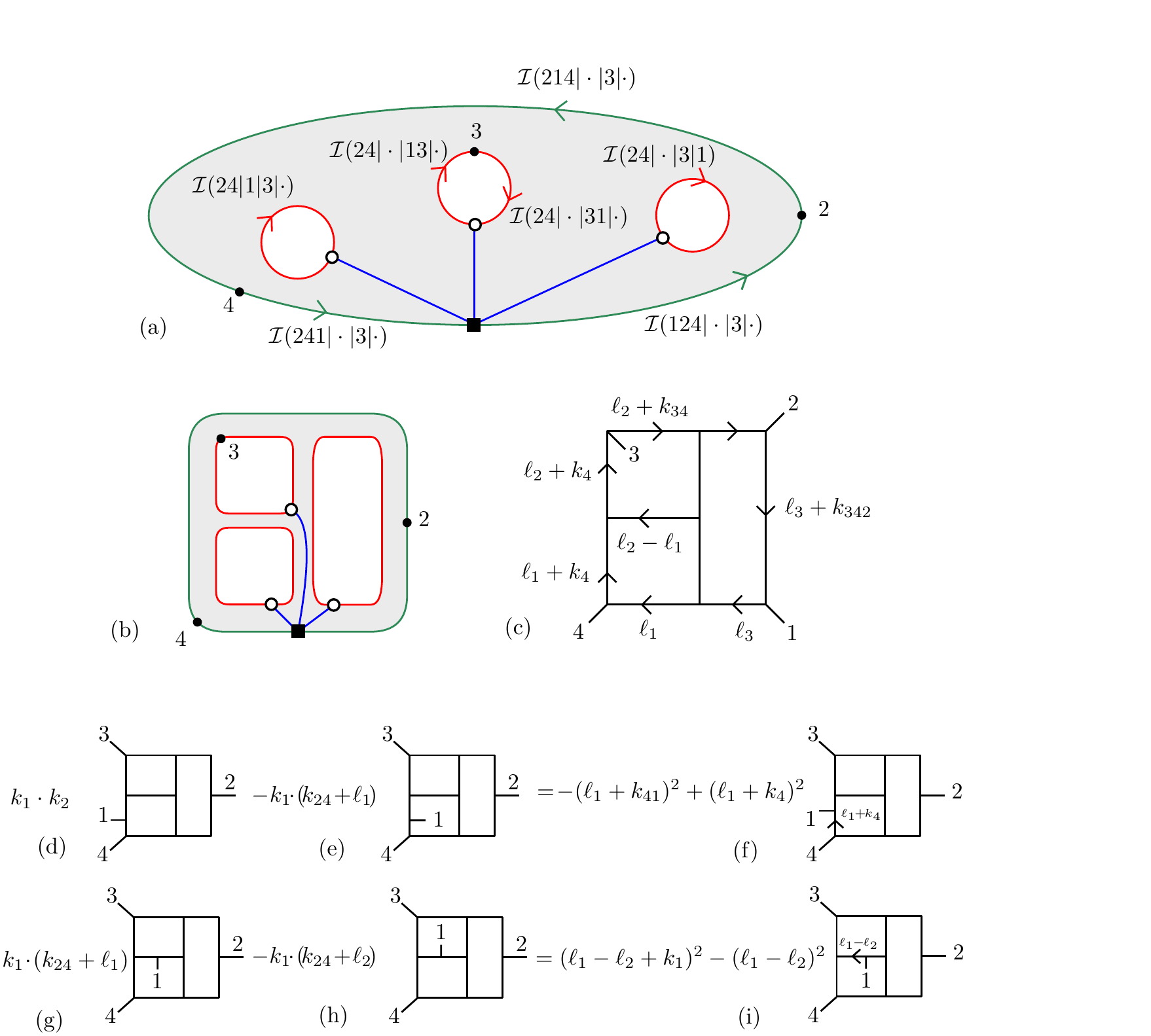}
	\caption{Proof that the accumulation of phases is a requirement for BCJ compatibility.
		(a) The 3-loop graph and monodromies we consider. (b) A pictorial representation of the Mercedes degeneration. The moduli space at three loops encode the other topology as well, the ladder graph, so this is only one type of degeneration that enters eq.~\eqref{eq:3-loop-monodromy}. (c) A particular graph appearing the field theory limit of $I(12|\cdot|3|\cdot)$. Note that, because it is the amplitude from which we do the analytic continuation (it defines the physical branch), this graph does not enter the field theory monodromies coming from eq.~\eqref{eq:3-loop-monodromy} at order $\alpha'$ (again, see details in \cite{Tourkine:2016bak,Ochirov:2017jby,Tourkine:2019ukp}). (d)$\to$(i) Other graphs entering various terms of the monodromy relations of eq. ~\eqref{eq:3-loop-monodromy}.}
	\label{fig:3loop}
\end{figure}

As a consistency check on the relations \eqref{eq:g_amp_rel}, we illustrate below how the accumulation of phases is a necessary phenomenon for compatibility with field theory BCJ relations for the non-planar graphs. The phase accumulation cannot be seen from a single BCJ triplet, so one is required to look globally at the BCJ properties of the field theory limit, making the following check somewhat lengthy. In the field theory limit of terms in~\eqref{eq:g_amp_rel}, BCJ identities are created by propagator cancellation induced by \textit{differences of phases}, such as
\begin{equation}
  \label{eq:BCJ-ex_1}
 2 k_1\cdot(\ell_J + k_i)-2k_1\cdot\ell_J = (\ell_J+k_1+k_i)^2-(\ell_J+k_i)^2
\end{equation}
or 
\begin{equation}
  \label{eq:BCJ-ex_2}
 2 k_1\cdot\ell_J = (\ell_J+k_1)^2-\ell_J^2\,.
\end{equation}
The constant shifts caused by the phase accumulation, such as replacing $k_1\cdot(\ell_J + k_i)$ by $k_1\cdot(\ell_J + k_i)+\Delta$, are thus invisible in these differences. Only when comparing across various graphs one can see the necessity of phase accumulation.\footnote{The reader is referred to \cite{Tourkine:2016bak,Ochirov:2017jby} and especially \cite{Tourkine:2019ukp} for a review of the BCJ-compatibility mechanism.}

Let us now proceed to give a consistency check on the phase accumulation. Consider the identity below in a maximally supersymmetric gauge theory in $D=4$
\begin{multline}
  \label{eq:3-loop-monodromy}
  \mathcal{I}(124|\cdot|3|\cdot)+
  e^{i\pi k_1\cdot k_2}\mathcal{I}(214|\cdot|3|\cdot)+
  e^{i\pi k_1\cdot k_{24}}\mathcal{I}(241|\cdot|3|\cdot)+
  e^{i\pi k_1\cdot (k_{24}+\ell_1)}\mathcal{I}(24|1|3|\cdot)+\\
  e^{i\pi k_1\cdot (k_{24}+\ell_2)}\mathcal{I}(24|\cdot|13|\cdot)+
  e^{i\pi k_1\cdot \ell_2}\mathcal{I}(24|\cdot|31|\cdot)+
  e^{i\pi k_1\cdot \ell_3}\mathcal{I}(24|\cdot|3|1)=\sum_{I=1}^3 \mathcal{J}_I(\ell )-\mathcal{J}_I(\ell +k_1).
\end{multline}
In Figure~\ref{fig:3loop}, graph (d), we see that, although $k_2$ is not related to the first loop, the phase $k_1\cdot k_2$ needs to be carried over in graph (e), for otherwise the phase from (d) $k_1\cdot k_{2}$ is not canceled and the inverse propagators of graph (f) are not neatly reconstructed. Likewise for particles 4 and 2 in graph (h); the phase needs to be carried over, otherwise the phase from the first diagram $k_1\cdot k_{42}$ is not canceled and the inverse propagators of graph (i) are not reconstructed either.

This leads to the conclusion that the phase accumulation is consistent with the BCJ compatibility mechanism in the non-planar sector and strengthens the necessity of this accumulation.

\section{Discussion}
\label{sec:discussion}

\subsection{Counting the number of independent integrands}

Determining the number of independent string integrands at genus one and higher is more subtle than at tree level. Firstly, the monodromy relations involve contours along the halved $A$-cycles which are not physical contours from the open string theory's perspective. Secondly, if we do not allow shifts of the loop momenta, which mean we really work at fixed loop momentum, the contours running along the $A$-cycles which differ by a translation along a $B$-cycle are not identified, and therefore they should be counted as independent. For instance, although we know that $\mathcal{J}_{+}(\ell)=\mathcal{J}_-(\ell+k_1)$, our set-up seems to indicate that these cycles should be counted differently, and it is not clear what this imply for the open string.

\begin{figure}[]
  \centering
  \includegraphics{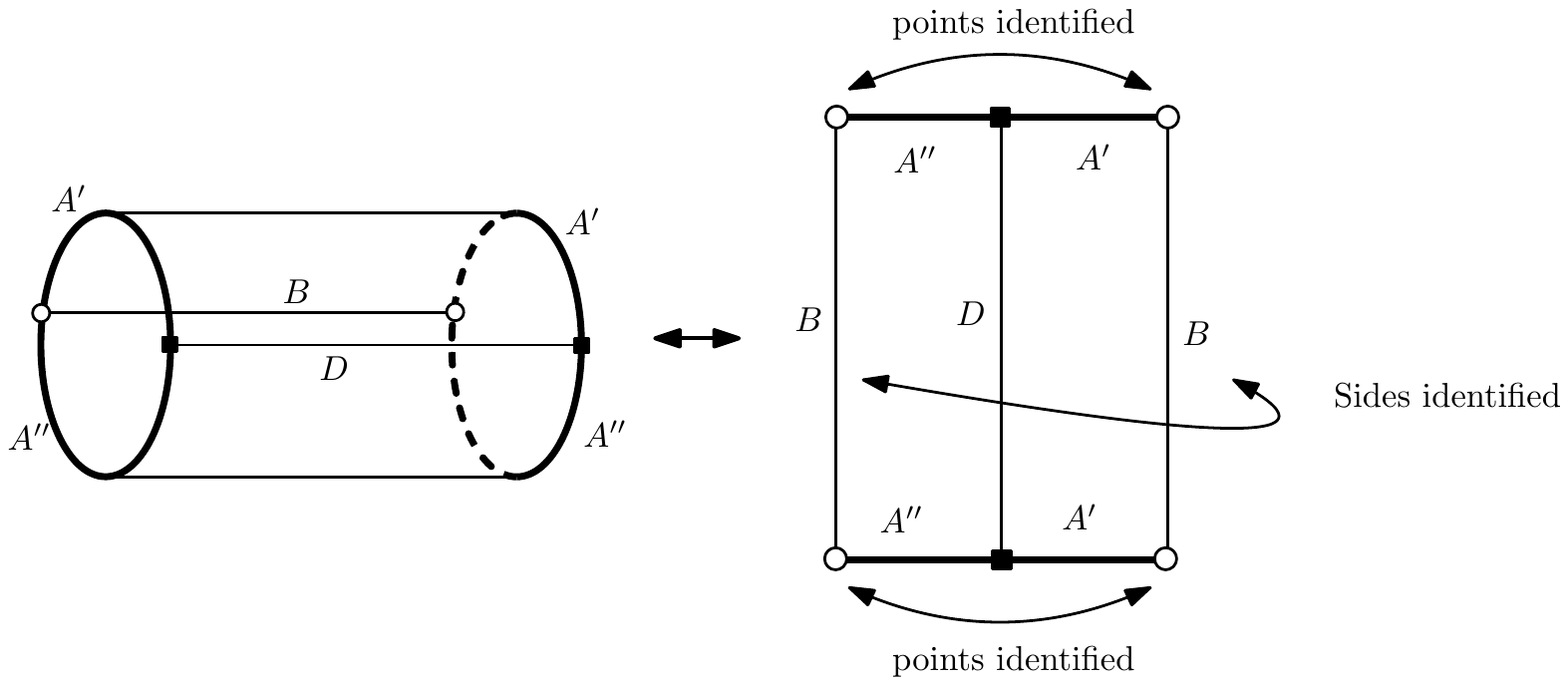}
  \caption{Example of computation of $\chi(\tilde \Sigma_{g,0})$ with $g=1$. Using the definition $\chi=F-E+V$ we find $\chi(\tilde \Sigma_{1,0})=0$, just like for the normal torus. At higher genus, this shows us that the points (white dots) on each side of the reflection axis $D$ must be identified in Figure~\ref{fig:gcut_2}, while the top and bottom points (black boxes) are not; the $B_i$ cycles must also be identified, but the $A'_i$ or $A''_i$ must not. This gives the result of eq.~\eqref{EC_cut_surface}.}
  \label{fig:chi}
\end{figure}

If we apply the same reasoning as in Section~\ref{sec:review}, we would be tempted to relate the number of twisted cycles to the Euler characteristics of the configuration space of the punctures on the cut surfaces, $\tilde F_{g,n}$. The later can be computed recursively using the forgetful map $\tilde F_{g,n}\to \tilde F_{g,n-1}$, whose fiber is the cut surface with one point removed, $\tilde \Sigma_{g,n-1}$. By identifying the points as illustrated in Figure~\ref{fig:chi}, we find, by a simple triangulation, that
\begin{align}\label{EC_cut_surface}
\chi(\tilde\Sigma_{g,n-1}) = \chi(\tilde\Sigma_g) - (n{-}1) = -(2g{-}2+(n{-}1)).
\end{align}
We can then compute $\chi(\tilde F_{g,n})$ multiplicatively as before,
\begin{align}
\chi(\tilde F_{g,n}) &= \chi(\tilde F_{g,n-1})\,\chi(\tilde\Sigma_{g,n-1})\nn\\
&= \textstyle\prod_{k=\max(1,4{-}2g)}^{n} \chi(\tilde\Sigma_{g,k-1}),
\end{align}
to obtain
\begin{equation}\label{EC_1}
\chi(\tilde F_{g,n}) = \begin{dcases}
(-1)^{n-3} (n{-}3)! \qquad & \text{for } g=0, n\geq 3,\\
(-1)^{n-1} (n{-}1)! \qquad & \text{for } g=1, n\geq 1,\\
(-1)^{n} (2g{+}n{-}3)!/(2g{-}3)! \qquad & \text{for } g \geq 2,
\end{dcases}
\end{equation}

In the absence of extra symmetries in the construction and under some genericity assumptions this would the number of independent twisted cycles. In our context it only provides an upper bound; and we can argue that the number of open string twisted cycles will be much smaller than this.
The key here is that to obtain open string amplitudes one should apply a $\mathbb{Z}_2$ involution that folds the graph of Figure~\ref{fig:gcut_2} along the D axis. Not every closed surface admits such an involution, it is only available for particular values of the modular parameters $\Omega_{IJ}$. Thus it is a further symmetry that is somewhat hidden in the way we set up our framework.
It is, therefore, tempting to think that the number of cycles should instead be related to the Euler characteristics of the configuration space of the annulus with $g$ holes and $n$ punctures on these boundaries, $G_{g,n}$:
\begin{equation}
  \label{eq:chi-annulus}
  \chi(G_{g,n}) = \frac{(g+n-2)!}{g!},
\end{equation}
for $g \geq 1$.

This number would have to be compared to the number total number of open string configurations. The later can be evaluated to be $(n+g)!/g!$ at $g$-loops. It is obtained by simply enumerating all the ways to distribute $n$ points on the $g+1$ boundaries of our cut surface, folded in half.

Therefore, the main problem of the previous Euler characteristics counting is the factor of $2$ in front of $g$ in the factorial. One can see where it comes from by looking at a configuration where all the particles are on the $D$ cycle. Then, one can draw twisted cycles between each of the black boxes of Figure~\ref{fig:gcut_2}, and because we do not identify the $A'$ and $A^{\prime\prime}$-cycles on each side, they all are taken as different cycles, given a total contribution of $2g$. The $\mathbb{Z}_2$ involution would precisely remedy this question by identifying these.

At this stage, one can only speculate about the analogue of the above discussion after integration of the loop momentum. It is however interesting to note that the Euler characteristics computed in \eqref{EC_1} are exactly those of the usual configuration space of points on Riemann surfaces (without cuts), and match the result of \cite{Harer1986}, and hence seems to relate to fully doubly periodic objects. This gives the hope that relations between \textit{amplitudes} may exist. Such a basis might in principle be smaller, for example, using additional integration-by-parts identities in the loop momentum space. Recently, a geometric approach to this problem based on twisted cohomologies was formulated in \cite{Mastrolia:2018uzb,Frellesvig:2019kgj,Frellesvig:2019uqt}. Analyzing relations between generic string theory amplitudes will likely involve a synthesis of these two approaches.

\subsection{Field-theory limit}

The all loop relations~\eqref{eq:g_amp_rel} give two relations in the field theory limit:\footnote{This is a different limit than the one discussed in the previous section. The expansion mentioned here is valid only at fixed $\tau$; the limit is taken \textit{outside} of the $\tau$ integral and the leading contribution are field theory Feynman diagrams, their possible UV divergences and stringy counterterms. This is well-known material from the study of the $\alpha'\to 0$ limit of string theory, see for instance \cite{Tourkine:2013rda,Vanhove:2014wqa,Amini:2015czm} and references therein.} one at $\mathcal{O}(1)$ and another at $\mathcal{O}(\alpha')$. It was observed in \cite{Tourkine:2016bak} that the former can be upgraded to a relation between loop \textit{amplitudes} since the phases only contribute $1$, and the loop momentum can be integrated out so that contributions from the $\J$ cycles cancel. At order $\mathcal{O}(\alpha')$ this is no longer possible; the relations depend on the phases (they are the analogous of the tree-level BCJ relations) and hold for field theory loop \textit{integrands}. Moreover, they receive contributions from the $\J$ cycles. It has been observed previously, in particular in \cite{Tourkine:2019ukp,Ochirov:2017jby}, that the field theory limit of the $\J$ cycles give rise to contact terms in the integrand representation, which act as correction terms that cure the loop-momentum shifting ambiguities which arise when performing BCJ moves (Jacobi identities) around the loop. This is a well-characterized problem in the BCJ literature; its solution could provide a better understanding on how to parametrize the anzatses used to compute supergravity integrands at very high loop orders \cite{Bern:2013qca,Bern:2014sna,Bern:2017yxu}, up to five \cite{Bern:2018jmv}.

Existing hints of the presence of monodromies in the field theory limit were already observed in the past. The works  \cite{Boels:2011tp,Boels:2011mn,Du:2012mt} at one-loop and \cite{Feng:2011fja,Chiodaroli:2017ngp} at higher loop where observed to be consistent with the stringy monodromy relations (those concerning planar amplitudes or sectors which were not affected by the discrepancies discussed above).
We leave a more detailed analysis of this aspect to future work, but we would like to emphasize that the calculations presented in this paper seem to connect in a rather deep way the twisted homology of string theory to the BCJ representations in field theory. This provides yet another example of the fact that the BCJ representation must carry some mysterious, but essential feature of field theory, yet to be discovered.

\subsection{Towards KLT relations}

At tree-level, KLT relations \cite{Kawai:1985xq} have a geometric interpretation in terms of intersection theory of twisted cycles on the moduli space ${\cal M}_{0,n}$ \cite{Mizera:2017cqs,Mizera:2019gea}. To be more precise, they can be understood as an insertion of identity in the space of twisted cycles into the closed-string amplitude, which decomposes it into two copies of open-string amplitudes. Thus the KLT matrix, given by intersection numbers of those cycles, measures non-orthogonality between the two bases of contours used for left- and right-movers. Therefore in generalizations of KLT relations to higher-genus amplitudes it is crucial to identify the basis of the relevant contours. In this paper we took first steps towards doing so.

One of our key results is that the basis will necessarily have to involve both physical and unphysical cycles. A consequence of this fact is that the corresponding KLT relations at finite $\alpha'$ will decompose a closed-string loop integrand in terms of quadratic combinations of not only open-string loop integrands, but also integrals over such unphysical cycles, at least in the present formulation. This fact further motives the study of the unphysical cycles and in particular how they contribute to the $\alpha' \to 0$ limit. We expect that these additional contributions produce diagrams with contact terms and would be related to similar construction in the loop-level double copy literature~\cite{Bern:2017yxu}.

It would be particularly interesting to study the relations to the recently-found double-copy structures at genus-one \cite{Mafra:2017ioj,Broedel:2018izr,Mafra:2018nla} in this context.

\appendix

\acknowledgments

We thank Stephan Stieberger and Pierre Vanhove for clarifying discussion about their work \cite{Hohenegger:2017kqy,Tourkine:2016bak,Ochirov:2017jby}.
S.M. and E.C. thank Dmitry Fuchs and Albert Schwarz for work on related topics. P.T. thanks Julio Parra Martinez and Henrik Johansson for discussions on the BCJ labeling question in particular, which motivate part of this research. P.T. thanks CERN for hospitality where part of this work was being carried.

S.M. gratefully acknowledges the funding provided by Carl P. Feinberg. This research is supported in part by U.S. Department of Energy grant DE-SC0009999 and by funds provided by the University of California. Research at Perimeter Institute is supported in part by the Government of Canada through the Department of Innovation, Science and Economic Development Canada and by the Province of Ontario through the Ministry of Economic Development, Job Creation and Trade.

\section{Properties of the prime form}
\label{prime_app}

In this appendix we review some basic properties of the prime form needed to define Green's functions on Riemann surfaces. We omit the dependence on the surface moduli $\Omega_{IJ} = \int_{B_I} \omega_J$ (where $\omega_J$ are Abelian differentials) from the notation as they always remain fixed in our computations.

On a Riemann surface the prime form $E(z,w)$ is the unique bi-holomorphic $(-\frac{1}{2},0)\times(-\frac{1}{2},0)$ form with a simple zero at $z=w$, that is, for $z$ sufficiently close to $w$
\begin{equation}
  \label{eq:E-local}
  E(z,w) = \frac{z-w}{\sqrt{dz}\sqrt{dw}}+ O(z-w)^3.
\end{equation}
It follows that the universal part of a genus-$g$ string integrand, which is given by
\begin{equation}
 \prod_{1 \leq i<j \leq n} E(z_i,z_j)^{k_i\cdot k_j},
\end{equation}
has the same local monodromies as the tree-level Koba--Nielsen factor $\prod_{1 \leq i<j \leq n} (z_i-z_j)^{k_i\cdot k_j}$.
The prime form is not double-periodic; it is only periodic in the $A_I$-cycle directions, but  when $z$ is moved along a $B_I$-cycle it changes by a phase
\begin{equation}\label{prime_period_compact}
  E(z+m_I A_I+n_I B_I,w) = \prod_{I=1}^{g}(-1)^{m_I + n_I}\exp\left(-i\pi\sum_{I=1}^{g} n_I \Omega_{II}-2\pi i\sum_{I=1}^{g} n_I \textstyle\int_z^w\omega_I\right) E(z,w),
\end{equation}
where $\omega_I$ is a basis of Abelian differentials and $\Omega_{II}$ are diagonal entries of the period matrix for $I=1,2,\ldots,g$. Note that here, we disagree with the standard textbook by Fay~\cite{fay1973theta}, eq.~(20). The prime form should have a monodromy along the A cycles, as quoted in \cite{DHoker:1988pdl}. An easy way to see this is that, if we take one of the A cycles to pinch, it has alternatively a description along a very long tube, for which $E(z)\sim \sin(\pi z)/\pi$, which picks up a sign when $z\to z+1$, similarly to what happens at one-loop for the function $\theta_1(z,\tau)$. When $\tau\to i \infty$, we have $\ln(\theta_1(z,\tau)) = 1/\pi\ln(\sin(\pi z))+O(q)+c(\tau)$ where $c(\tau)$ is an irrelevant constant.

An explicit construction of the prime form can be given using the Riemann theta functions, defined as
\begin{equation}
\theta\ab ( \zeta| \Omega) :=
\sum_{ n\in\mathbb{Z}^g} e^{i\pi ({n}+{\beta})\cdot  \Omega ({n}+{\beta})} e^{2 i
	\pi ({n}+{\beta})\cdot ({\zeta}+{\alpha})},
\label{eq:thetachar}
\end{equation}
where $\ab= \nu \in (\mathbb{Z}/\mathbb{Z}_2)^{2g}$ is a theta
characteristic and $\Omega$ the period matrix. A characteristic $\ab$ is called even or odd according to
the parity in $z$ of the corresponding theta function when transported along $A_I$- and $B_I$-cycles. A theta function is called even or odd according to the parity of the scalar product $\beta\cdot\alpha$ mod
2. There are $2^{g-1}(2^g+1)$ even theta characteristics and
$2^{g-1}(2^g-1)$ odd ones, respectively. Let $\gamma,\delta$ be elements of $\frac 12(\mathbb{Z}/2\mathbb{Z})^2$. Then theta functions transforms as
\begin{equation}
\label{eq:half-period-Riemann}
\theta\ab(\zeta+\gamma+\Omega\delta) = e^{-i\pi
	\delta\cdot\Omega\delta-2i\pi \delta\cdot(\zeta+\alpha+\gamma)} \theta{\left[ \begin{smallmatrix} {\beta+\delta}\\
	{\alpha+\gamma} \end{smallmatrix} \right]}(\zeta|\Omega).
\end{equation}

Using~\eqref{eq:thetachar}, the prime form is defined on the universal covering of the surface by
\begin{equation}
E(z,w) :=
\frac{\theta[ \nu](\int_x^y
	(\omega_1,...,\omega_g)|{\Omega})}{h_{ \nu}(z) h_{ \nu}(w)}\,\in 
\mathbb{C}\,,
\label{e:primeform}
\end{equation}
where $h_{ \nu}(z)^2 = \sum_{I=1}^{g} \omega_I(z) \partial_I \theta[ \nu]
(0|\Omega)$ are half-differentials (sections of the square-root of the
canonical bundle). It is independent of the spin structure chosen
to define it.

The transformation properties in~\eqref{prime_period_compact} can be used to show that there is an extra phase that the prime form picks up when $z$ and $w$ are the position of particles on different boundaries, compared to when they are on the same boundary. In the former, the phase on the right-hand side has an integration path which runs over an extra half $A_I$-cycle, $2i\pi\int_{A_I/2}\omega_J = i\pi\delta_{IJ}$, which produces an extra sign when compared to the situation when both particles are on the same boundary. At genus one this boils down to the well known fact that $\vartheta_1$ and $\vartheta_2$ pick up different phases under a translation by the modular parameter $\tau$. This property is behind the fact that the bulk contributions $\J$ cannot in general be cancelled by a shift in loop momentum, only in the planar case where the phases cancel by momentum conservation.

\bibliographystyle{jhep}
\bibliography{biblio}

\end{document}